\documentclass[onecolumn,showpacs,preprintnumbers,elsart]{revtex4}
\usepackage{amsmath}
\usepackage{mathrsfs}
\usepackage{graphicx}
\usepackage{dcolumn}
\usepackage{bm}
\usepackage{color}
\usepackage[center]{subfigure}

\begin{document}

\title{Discrete solitons in self-defocusing systems with $\mathcal{PT}$%
-symmetric defects}
\author{Zhiqiang Chen$^{1}$}
\author{Jiasheng Huang$^{1}$}
\author{Jinglei Chai$^{1}$}
\author{Xiangyu Zhang$^{1,2}$}
\author{Yongyao Li$^{1}$}
\email{yongyaoli@gmail.com}
\author{Boris A. Malomed$^{3}$}
\affiliation{$^{1}$Department of Applied Physics, South China Agricultural University,
Guangzhou 510642, China \\
$^{2}$ Department of Electrical and Computer Engineering, Duke University,
Durham, North Carolina 27708, USA\\
$^{3}$ Department of Physical Electronics, School of Electrical Engineering,
Faculty of Engineering, Tel Aviv University, Tel Aviv 69978, Israel.}

\begin{abstract}
We construct families of discrete solitons (DSs) in an array of
self-defocusing waveguides with an embedded $\mathcal{PT}$
(parity-time)-symmetric dimer, which is represented by a pair of waveguides
carrying mutually balanced gain and loss. Four types of states attached to
the embedded defect are found, namely, staggered and unstaggered bright
localized modes and gray or anti-gray DSs. Their existence and stability
regions expand with the increase of the strength of the coupling between the
dimer-forming sites. The existence of the gray and staggered bright DSs is
qualitatively explained by dint of the continuum limit. All the gray and
anti-gray DSs are stable (some of them are unstable if the dimer carries the
\textit{nonlinear} $\mathcal{PT}$ symmetry, represented by balanced
nonlinear gain and loss; in that case, the instability does not lead to a
blowup, but rather creates oscillatory dynamical states). The boundary
between the gray and anti-gray DSs is predicted in an approximate analytical
form.
\end{abstract}

\maketitle

%\email{yongyaoli@gmail.com}
%\email{stszjy@mail.sysu.edu.cn}

%\preprint{APS/123-QED}

% Force line breaks with \\



\section{Introduction}

Dynamics of discrete systems has been a subject of intensive studies in
diverse areas of physics, including dynamical lattices and long molecules,
optics, ultracold atomic gases, lattice QCD, etc. \cite{Aubry}-\cite{Kenig2}%
. In particular, it is well established that optical discrete solitons (DSs)
readily self-trap in nonlinear waveguiding arrays \cite%
{Christodoulides,Lederer}. In addition to their significance to fundamental
studies, DSs offer various possibilities for all-optical data-processing
applications; in particular, they can implement intelligent functional
operations, such as routing, blocking, logic functions and time-gating \cite%
{Christodoulides2}. Therefore, methods allowing one to control the
formation, mobility and interactions of DSs have been a subject of many
theoretical and experimental studies.

It is well known too that light confinement can be realized with the help of
various defects. Linear photonic defects can be created as localized
structures in photonic crystals \cite{5}, nanocavities \cite{6},
microresonators \cite{7} and quantum-dot settings \cite{8}. In particular,
defects have been designed to control DSs in arrayed waveguides \cite{Cao}-%
\cite{yongyao}. Nonlinear defects in photonic arrays have also been
elaborated, chiefly theoretically \cite{Molina-Ts}-\cite{Miro}.

Recently, attention has been drawn to defects formed by parity-time ($%
\mathcal{PT}$)-symmetric dimers, i.e., pairs of cores carrying mutually
balanced gain and loss, embedded into waveguide arrays \cite{Suchkov,Miro,we}
(related settings are represented by gain cores embedded into dissipative
lattices \cite{Ding1D}-\cite{Ding2D}), as well as continuum counterparts of
such systems, with the embedded dimer (alias a $\mathcal{PT}$-symmetric
dipole) represented by a combination of the delta function and its
derivative, in the real and imaginary parts, respectively \cite{Thawatchai}.
These lattice systems, which are governed by discrete nonlinear Schr\"{o}%
dinger (DNLS) equations corresponding to $\mathcal{PT}$-symmetric
non-Hermitian Hamiltonians \cite{Bender}, \cite{Muga}-\cite{undestr}, give
rise to entirely real propagation spectra, provided that the strength of the
gain and loss terms does not exceed a critical level, past which the $%
\mathcal{PT}$ symmetry suffers spontaneous breaking (a possibility of having
\textit{unbreakable} $\mathcal{PT}$ symmetry was recently reported in a
model incorporating self-defocusing nonlinearity with the local strength
growing fast enough from the center to periphery \cite{undestr}). Linear $%
\mathcal{PT}$ systems were realized experimentally in optics, by coupling
pumped and lossy waveguides \cite{Ruter}-\cite{Regensburger}. The simplest
version of $\mathcal{PT}$-symmetric nonlinear systems was elaborated
theoretically in the form of dimers with the onsite Kerr \cite{Barash}-\cite%
{KLi} or quadratic \cite{chi2} terms. A \textit{nonlinear} version of the $%
\mathcal{PT}$ symmetry, represented by the balanced nonlinear gain and loss,
was introduced too \cite{Kartash,Miro}.

Previous works on $\mathcal{PT}$-symmetric dimers embedded into lattices
were dealing with the self-focusing nonlinearity or linear lattices \cite%
{Suchkov,Miro,we}, while self-defocusing is also possible in photonics \cite%
{Staliunas,Lederer}. In this work, we introduce the system with a $\mathcal{%
PT}$-symmetric dimer embedded into a one-dimensional array of
self-defocusing waveguides. The system is described by a DNLS equation with
a defect representing the dimer. As a generalization, we also briefly
consider the dimer with the nonlinear $\mathcal{PT}$ symmetry. We find that
the system supports stable staggered and unstaggered localized modes (bright
DSs pinned to the defect), along with gray and anti-gray DSs (the latter
means a soliton featuring a local elevation on top of a flat background \cite%
{DFrantz}). Existence regions for them are found in a partly analytical
form, using the continuum limit of the discrete systems. The stability of
the DSs is investigated by means of numerical methods, \textit{viz}.,
calculation of eigenvalues for small perturbations, and direct simulations
of the underlying DNLS equation.

The paper is structured as follows. The models are introduced in section II.
Bright DSs (staggered and unstaggered ones) and gray and anti-gray DSs are
studied, respectively, in section III and IV (the latter section also
includes the consideration of DSs pinned to the defect with the nonlinear $%
\mathcal{PT}$ symmetry). The paper is concluded by section VI.

\section{The models}

\subsection{The system with the linear $\mathcal{PT}$ symmetry}

\begin{figure}[tbp]
\centering{\label{fig1a} \includegraphics[scale=0.4]{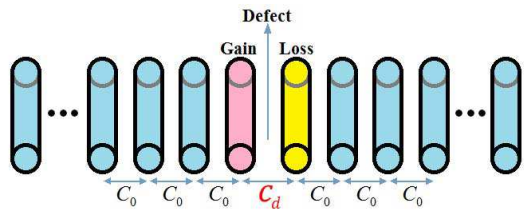}}
\caption{(Color online) The schematic of the nonlinear waveguide array with
the defect represented by the embedded $\mathcal{PT}$-symmetric dimer.}
\label{Model}
\end{figure}

The lattice system with the defect carrying the linear $\mathcal{PT}$
symmetry is based on the DNLS equation written as

\begin{equation}
i{\frac{du_{n}}{dz}}=-\left( C_{n-1,n}u_{n-1}+C_{n,n+1}u_{n+1}\right)
+|u_{n}|^{2}u_{n}+i\kappa _{n}u_{n},  \label{DNLS}
\end{equation}%
where $u_{n}$ is the amplitude of light in the $n$-th core of the arrayed
waveguide, $z$ is the propagation distance, $C_{n,n+1}$ and $\kappa _{n}$
are the coupling constant and gain-loss coefficient, respectively. As said
above, the array features the self-defocusing on-site nonlinearity and an
embedded defect, which is formed by the pair of sites with a tunable
strength, $C_{d}$, of the coupling between them, see Fig. \ref{Model}. The
two defect-forming sites carry mutually balanced linear gain and loss, which
is described by $\kappa $ and $-\kappa $ ($\kappa >0$).

Thus, coefficients $C_{n,n+1}$ and $\kappa _{n}$ in Eq. (\ref{DNLS}) are
defined as
\begin{equation}
C_{n,n+1}=%
\begin{cases}
C_{d} & \mathrm{at~~}n=-1, \\
C_{0} & \mathrm{at~~}n\neq -1,%
\end{cases}%
\quad \kappa _{n}=%
\begin{cases}
\kappa  & \mathrm{at~~}n=-1, \\
-\kappa  & \mathrm{at~~}n=0, \\
0 & \mathrm{elsewhere},%
\end{cases}
\label{kappa}
\end{equation}%
where $C_{0}$ is the inter-site coupling constant outside of the defect, and
$N$ is the size of the array. It is implied that $C_{d}/C_{0}>1$ and $C_{d}/C_{0}<1$ correspond to the distance
between the defect-forming sites which is, respectively, smaller or larger
than the separation between the sites outside of the defect. Propagating
modes are characterized by the total field power (alias norm of the
solution),
\begin{equation}
P=\sum_{n=-N/2}^{N/2-1}|u_{n}|^{2}.  \label{P}
\end{equation}%
Hereafter, we fix $C_{0}=1/2$ by means of obvious rescaling, and produce
numerical results for the system of size $N=128$, with $P$, $C_{d}$ and $%
\kappa $ treated as control parameters.

Stationary solutions to Eq. (\ref{DNLS}) with real propagation constant $%
-\mu $ are looked for as
\begin{equation}
u_{n}(z)=U_{n}e^{-i\mu z},  \label{mu}
\end{equation}%
where $U_{n}$ is the distribution of the local amplitudes. Stationary
solutions were found in the numerical form by means of the
imaginary-time-propagation method \cite{Chiofalo}, while real-time
simulations of Eq. (\ref{DNLS}) were carried out using the four-step
Runge-Kutta algorithm with the periodic boundary conditions.

Stability of the localized stationary modes was investigated numerically by
means of computing eigenvalues for small perturbations, and the results were
verified by means of direct simulations of the perturbed evolution in the
framework of Eq. (\ref{DNLS}). The perturbed solution was taken as
\begin{equation}
u_{n}=e^{-i\mu z}(U_{n}+w_{n}e^{i\lambda z}+v_{n}^{\ast }e^{-i\lambda ^{\ast
}z}),  \notag
\end{equation}%
where the asterisk stands for the complex conjugate. The substitution of
this expression into Eq. (\ref{DNLS}) and linearization leads to the
eigenvalue problem for the perturbation wavenumber, $\lambda \equiv \lambda
_{\mathrm{r}}+i\lambda _{\mathrm{i}}$, and the eigenmodes, $\left\{
w_{n},v_{n}\right\} $:
\begin{equation}
\left(
\begin{array}{cc}
C-\mu +2|U_{n}|^{2}+i\kappa _{n} & U_{n}^{2} \\
-\left( U_{n}^{\ast }\right) ^{2} & -C+\mu -2|U_{n}|^{2}+i\kappa _{n}%
\end{array}%
\right) \left(
\begin{array}{c}
w \\
v%
\end{array}%
\right) =\lambda \left(
\begin{array}{c}
w \\
v%
\end{array}%
\right) .  \label{eigen}
\end{equation}%
Solution $U_{n}$ is stable if all eigenvalues $\lambda $ are real.

\subsection{Generalization for the nonlinear $\mathcal{PT}$ symmetry}

The lattice with the embedded dimer featuring the nonlinear $\mathcal{PT}$
symmetry is described by the following version of the DNLS equation:%
\begin{equation}
i{\frac{du_{n}}{dz}}=-\left( C_{n-1,n}u_{n-1}+C_{n,n+1}u_{n+1}\right)
+(1+i\kappa _{n})|u_{n}|^{2}u_{n},  \label{DNLS_N}
\end{equation}%
where coefficients $C_{n}$ and $\kappa _{n}$ are again defined as\ per Eqs. (%
\ref{kappa}). In terms of the optical realization, the nonlinear gain may be
provided by a combination of the usual linear amplification and saturable
absorption, while the nonlinear loss is usually induced by resonant
two-photon absorption \cite{Agrawal,Staliunas}. A more general system,
including linear and nonlinear $\mathcal{PT}$-symmetric terms, is possible
too \cite{Miro}, but the corresponding analysis is rather cumbersome.

\section{Bright modes}

\subsection{Staggered bright discrete solitons}

The standard staggering transformation is introduced by replacing the
lattice field in Eq. (\ref{DNLS}) by
\begin{equation}
u_{n}(t)\equiv (-1)^{n}\tilde{u}_{n}^{\ast }(t),  \label{stag}
\end{equation}%
where the asterisk stands for complex conjugate \cite{PGK}. The substitution
reverses the sign of the nonlinearity in the respective equation for $\tilde{%
u}_{n}$ into self-focusing:%
\begin{equation}
i{\frac{d\tilde{u}_{n}}{dz}}=-\left( C_{n-1,n}\tilde{u}_{n-1}+C_{n,n+1}%
\tilde{u}_{n+1}\right) +(-1+i\kappa _{n})|\tilde{u}_{n}|^{2}\tilde{u}_{n},
\label{tilde}
\end{equation}%
hence it can support bright solitons pinned to the defect carrying the gain
and loss. This possibility may be clarified in an analytical form, by
considering a continuum counterpart of Eq. (\ref{tilde}), with discrete
coordinate $n$ replaced by a continuous one, $x$, and a local defect of the
coupling constant represented by term $\varepsilon \delta (x)\left\vert d%
\tilde{u}(x)/dx\right\vert ^{2}$ in the respective Hamiltonian density, with
$\varepsilon \sim C_{d}-C_{0}$, see Eq. (\ref{kappa}), where $\delta (x)$ is
the delta-function. With a localized shape of a bright soliton, $\tilde{u}_{%
\mathrm{sol}}(x-\xi )$, whose center is placed at $x=\xi $, this term gives
rise to the effective potential for the soliton,
\begin{equation}
U(\xi )=\varepsilon \left\vert \frac{d\tilde{u}_{\mathrm{sol}}(\xi )}{d\xi }%
\right\vert ^{2}.  \label{U}
\end{equation}%
In particular, the usual bright-soliton shape, $\tilde{u}_{\mathrm{bright}%
}=A~\mathrm{sech}\left( a\xi \right) $, with constants $A$ and $a$, Eq. (\ref%
{U}) yields
\begin{equation}
U_{\mathrm{bright}}(\xi )=\varepsilon A^{2}a^{2}\sinh ^{2}\left( a\xi
\right) \mathrm{sech}^{4}\left( a\xi \right) ,  \label{min}
\end{equation}%
which features a potential minimum at $\xi =0$ for $\varepsilon >0$ and $%
\varepsilon <0$, respectively. Thus, the defect is attractive at $%
\varepsilon >0$, and repulsive at $\varepsilon <0$. Incidentally, this
argument explains the fact, reported in Ref. \cite{we}, that, in the case of
$C_{d}<C_{0}$, the pinned mode in the discrete system with the self-focusing
nonlinearity present solely at the two central sites carrying the gain and
loss, the pinned mode exists above a finite threshold value of the total
power (\ref{P}). Indeed, in this case the defect repels the solitary mode,
which must be compensated by the attraction induced by the nonlinearity
concentrated at the central sites, while in the opposite case, $C_{d}>C_{0}$%
, there is no threshold.

\begin{figure}[tbp]
\centering\subfigure[] {\label{fig2a}
\includegraphics[scale=0.25]{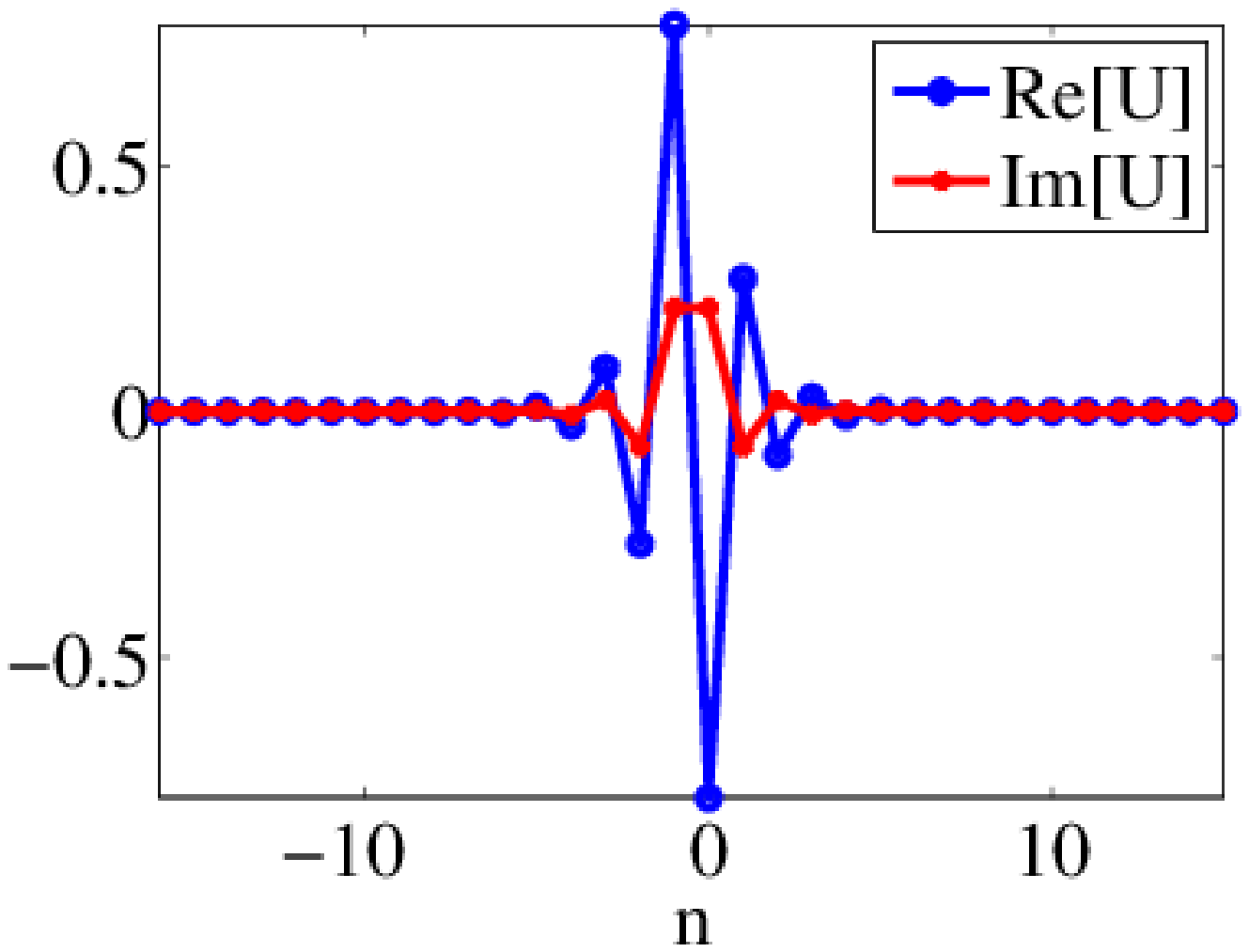}}%
\subfigure[] {\label{fig2b}
\includegraphics[scale=0.25]{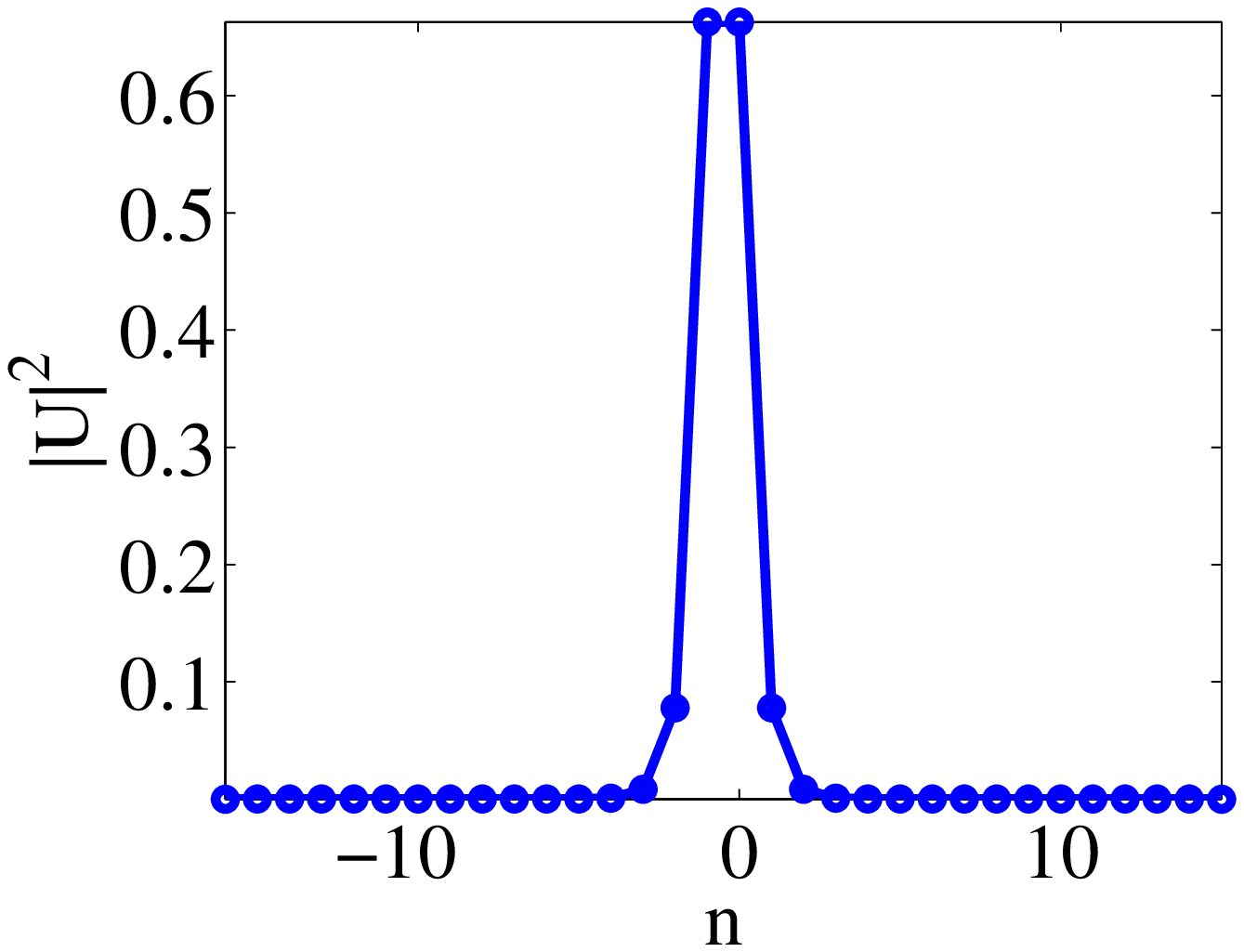}}
\subfigure[]{\label{fig2c}
\includegraphics[scale=0.25]{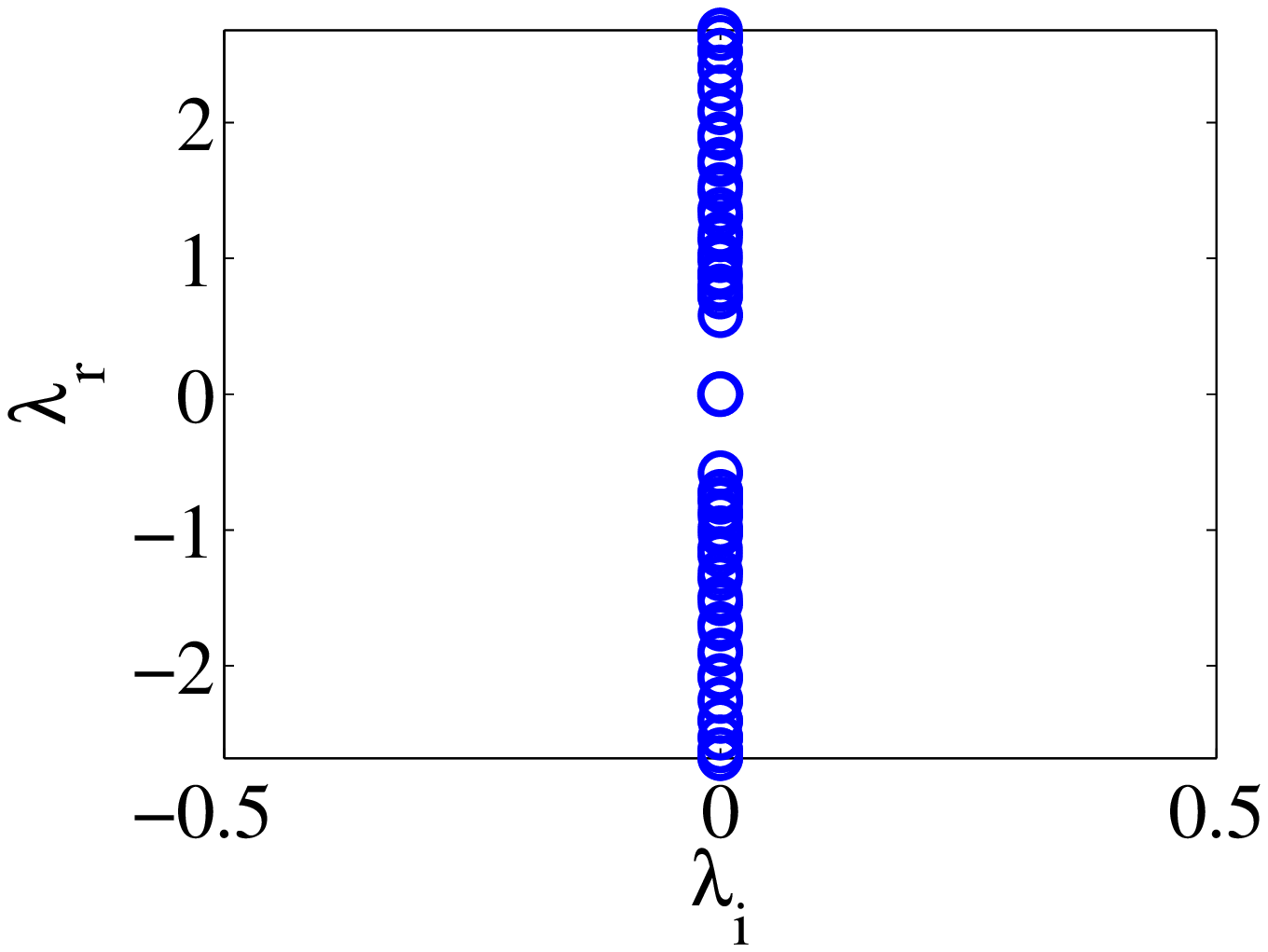}}
\subfigure[]{\label{fig2d}
\includegraphics[scale=0.25]{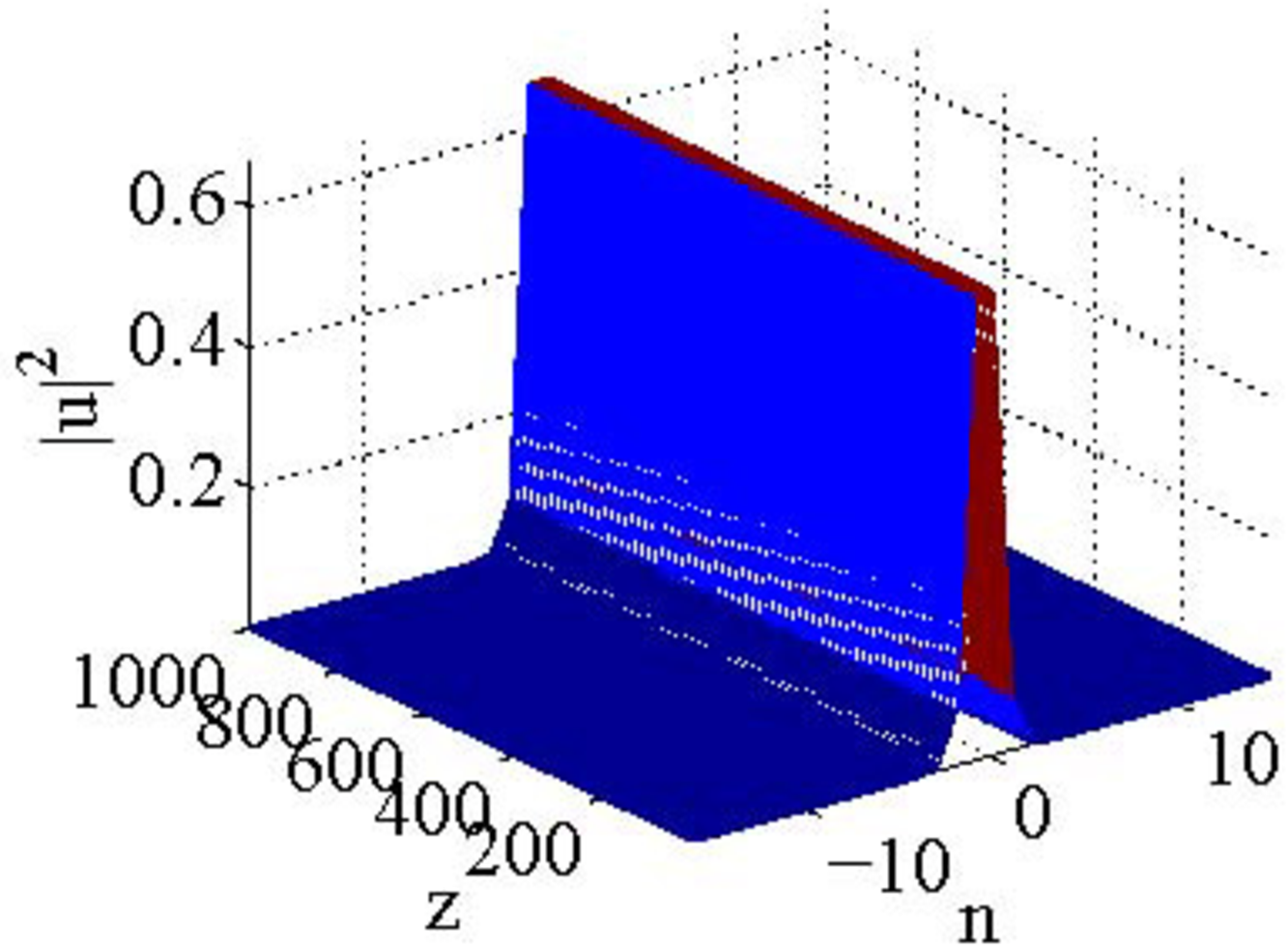}}
\caption{(Color online) A typical example of a stable bright staggered
soliton in the model based on Eq. (\protect\ref{DNLS}), with $(P,C_{d}/C_{0},%
\protect\kappa )=(1.5,2,0.5)$. (a) Real (blue) and imaginary (red) parts of
the solution. (b) The intensity profile of the soliton. (c) The spectrum of
stability eigenvalues (which demonstrates that this soliton is stable). (d)
Direct simulations of its perturbed evolution.}
\label{stablebright}
\end{figure}

\begin{figure}[tbp]
\centering\subfigure[] {\label{fig3a}
\includegraphics[scale=0.25]{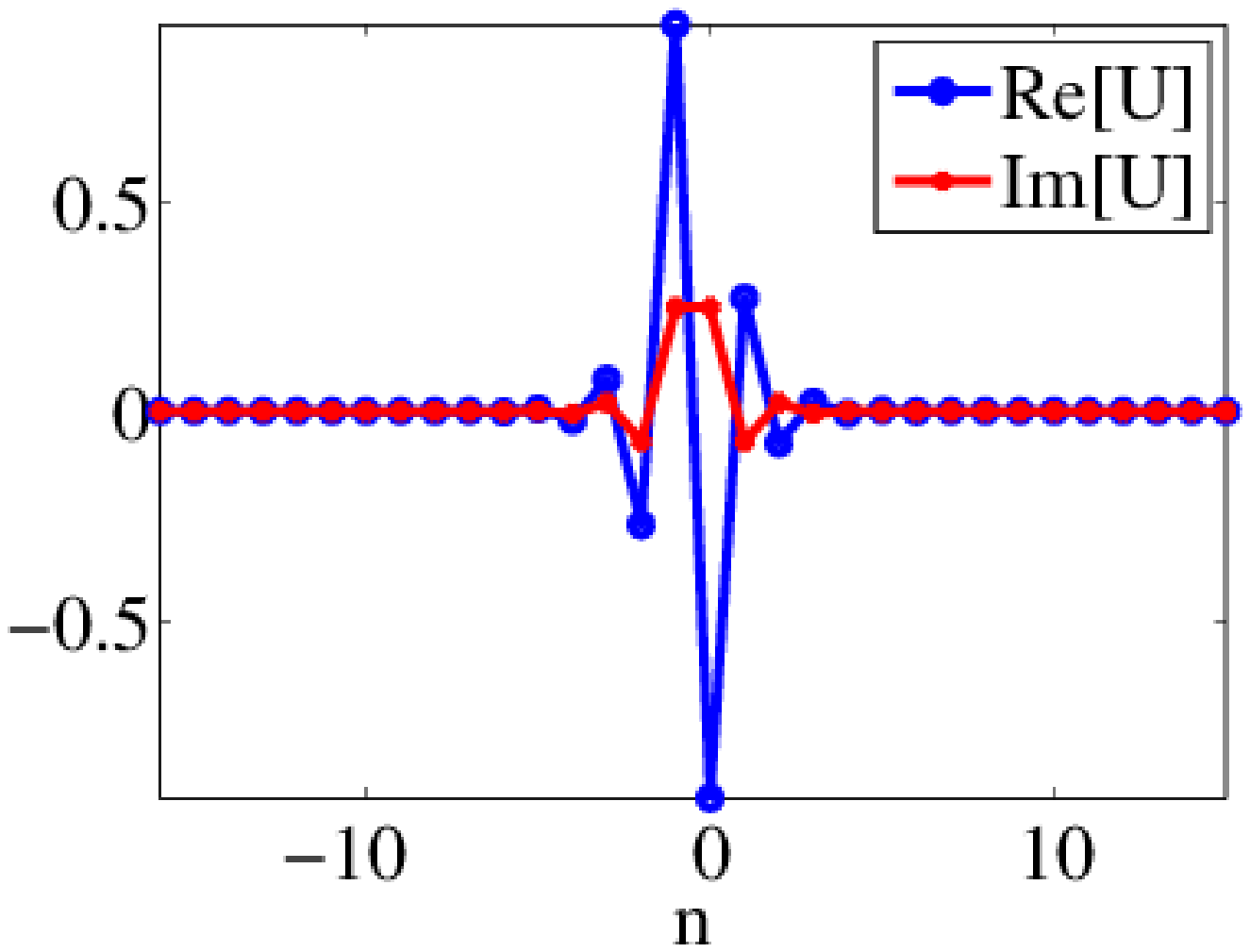}}%
\subfigure[] {\label{fig3b}
\includegraphics[scale=0.25]{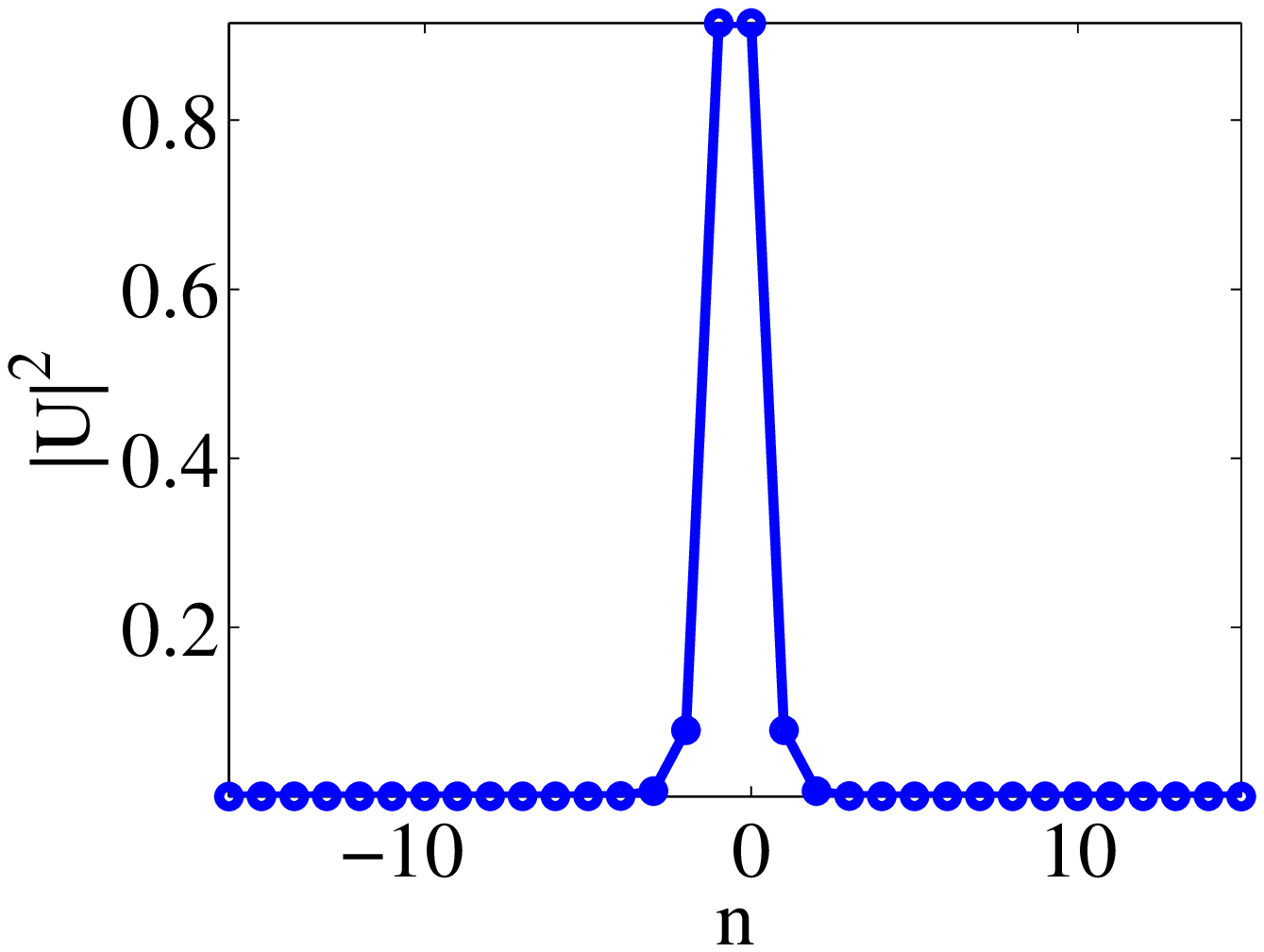}}
\subfigure[]{\label{fig3c}
\includegraphics[scale=0.25]{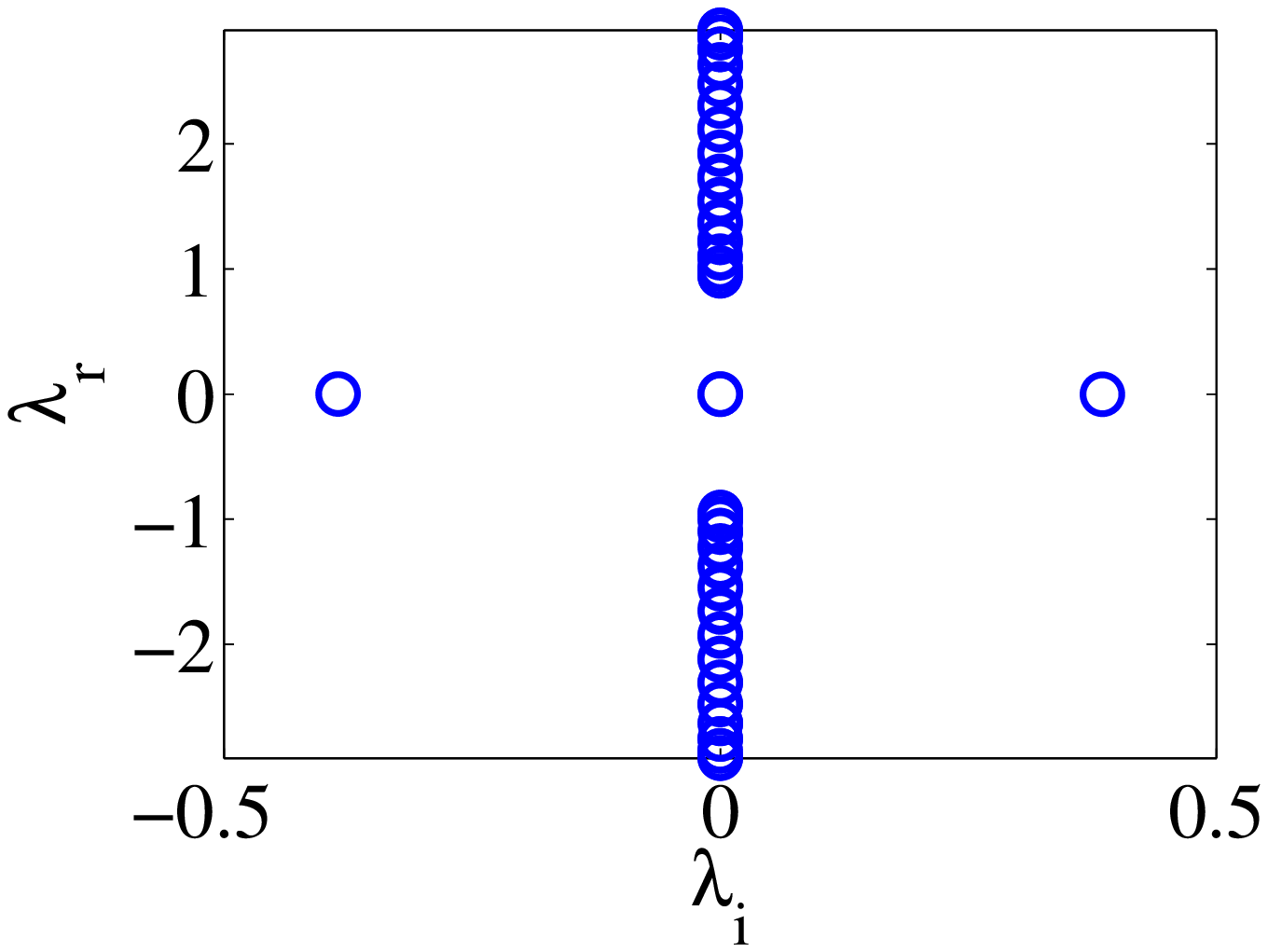}}
\subfigure[]{\label{fig3d}
\includegraphics[scale=0.25]{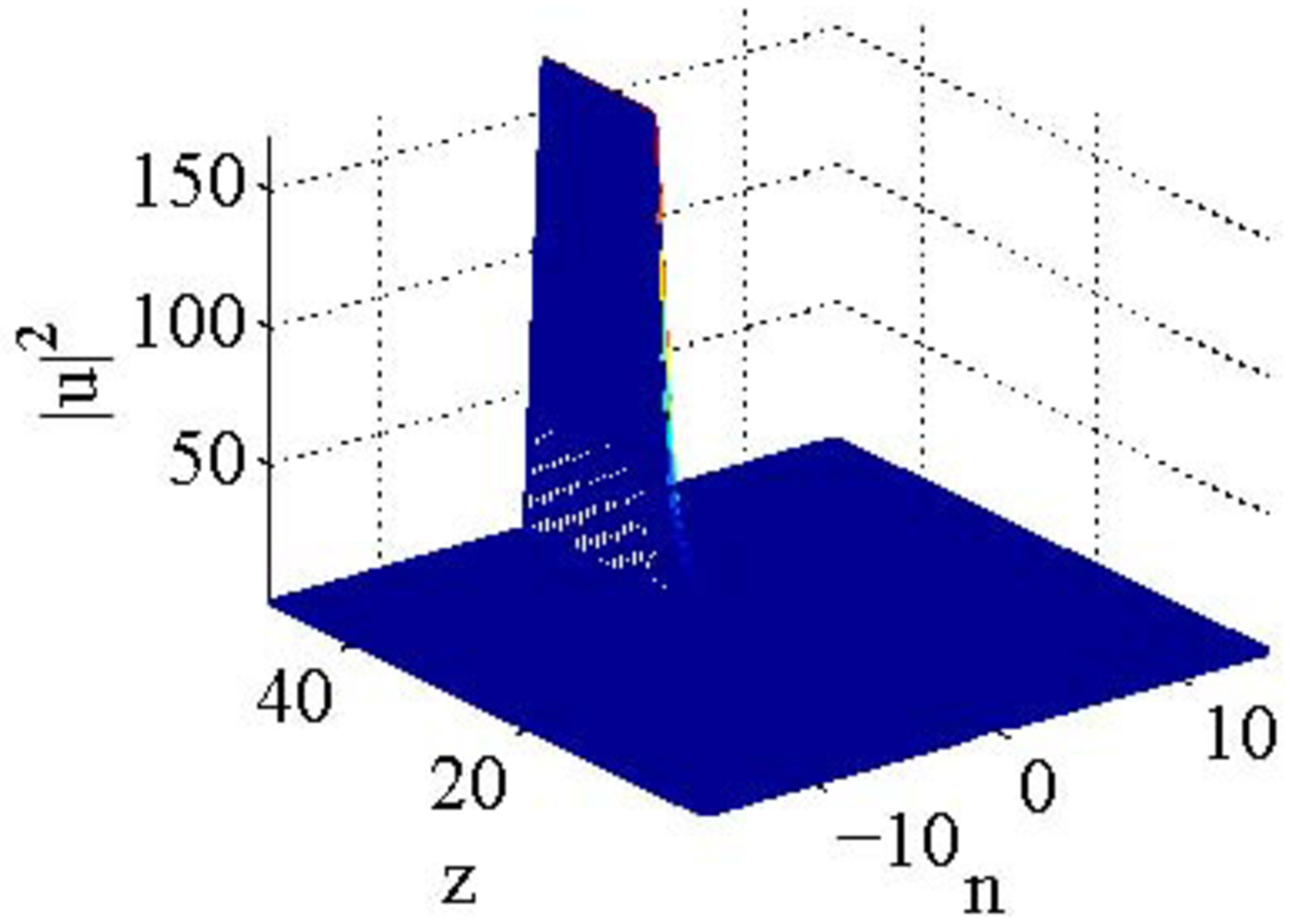}}
\caption{(Color online) The same as in the previous figure, but for an
unstable staggered soliton, with $(P,C_{d}/C_{0},\protect\kappa )=(2,2,0.5)$%
. }
\label{Unstablebright}
\end{figure}

Typical examples of stable and unstable staggered DSs, pinned to the $%
\mathcal{PT}$-symmetric defect, are displayed in Figs. \ref{stablebright}
and \ref{Unstablebright}, respectively. These figures clearly show that the
real and imaginary parts of the wave field are indeed staggered (the real
and imaginary parts are, severally, odd and even with respect to the
midpoint between $n=-1$ and $n=0$), while the intensity profile (the squared
absolute value of the field) does not exhibit any staggering. Direct
simulations demonstrate that the unstable DS undergoes a blowup under the
action of the defect.

\begin{figure}[tbp]
\centering\subfigure[] {\label{fig4a}
\includegraphics[scale=0.28]{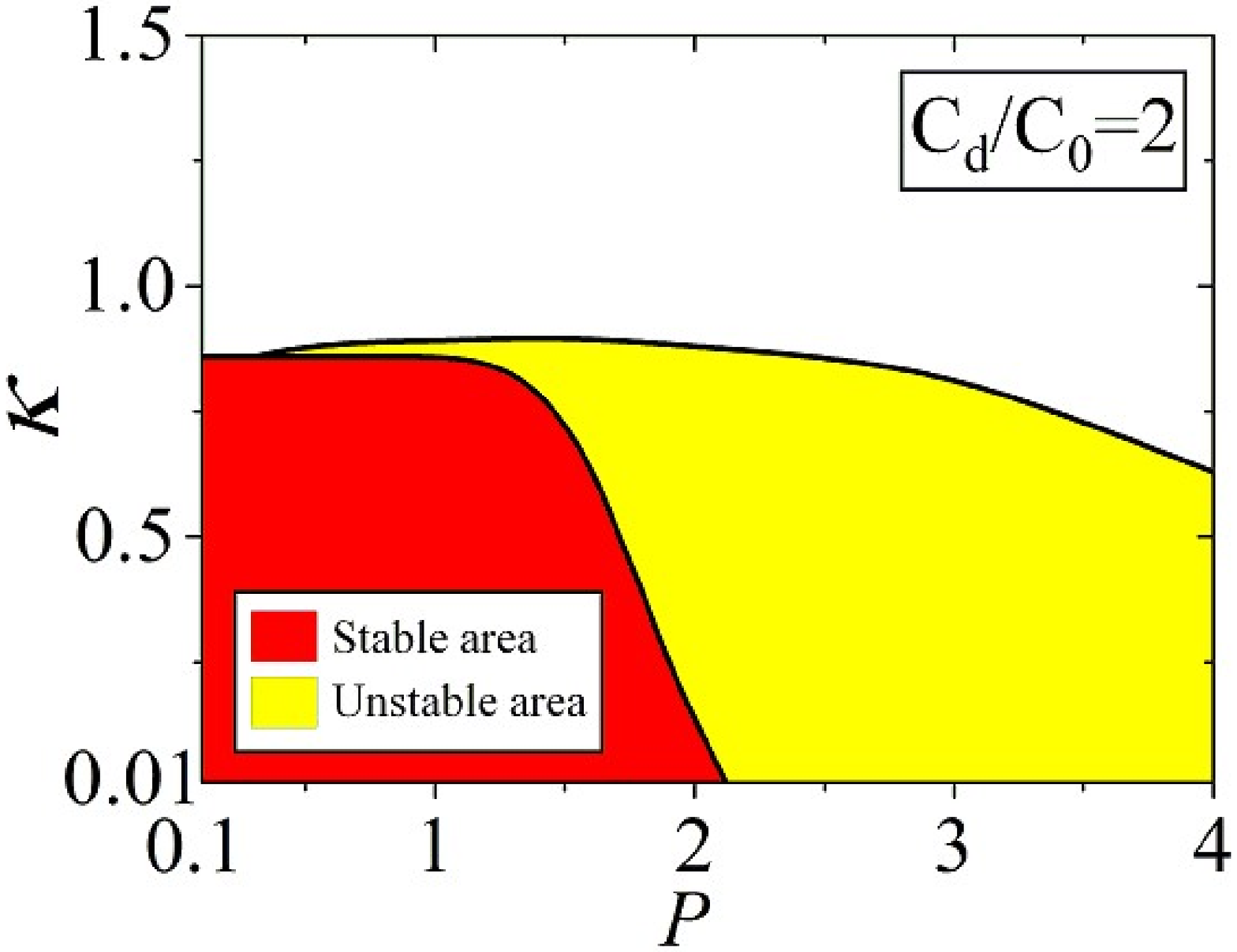}}%
\subfigure[] {\label{fig4b}
\includegraphics[scale=0.28]{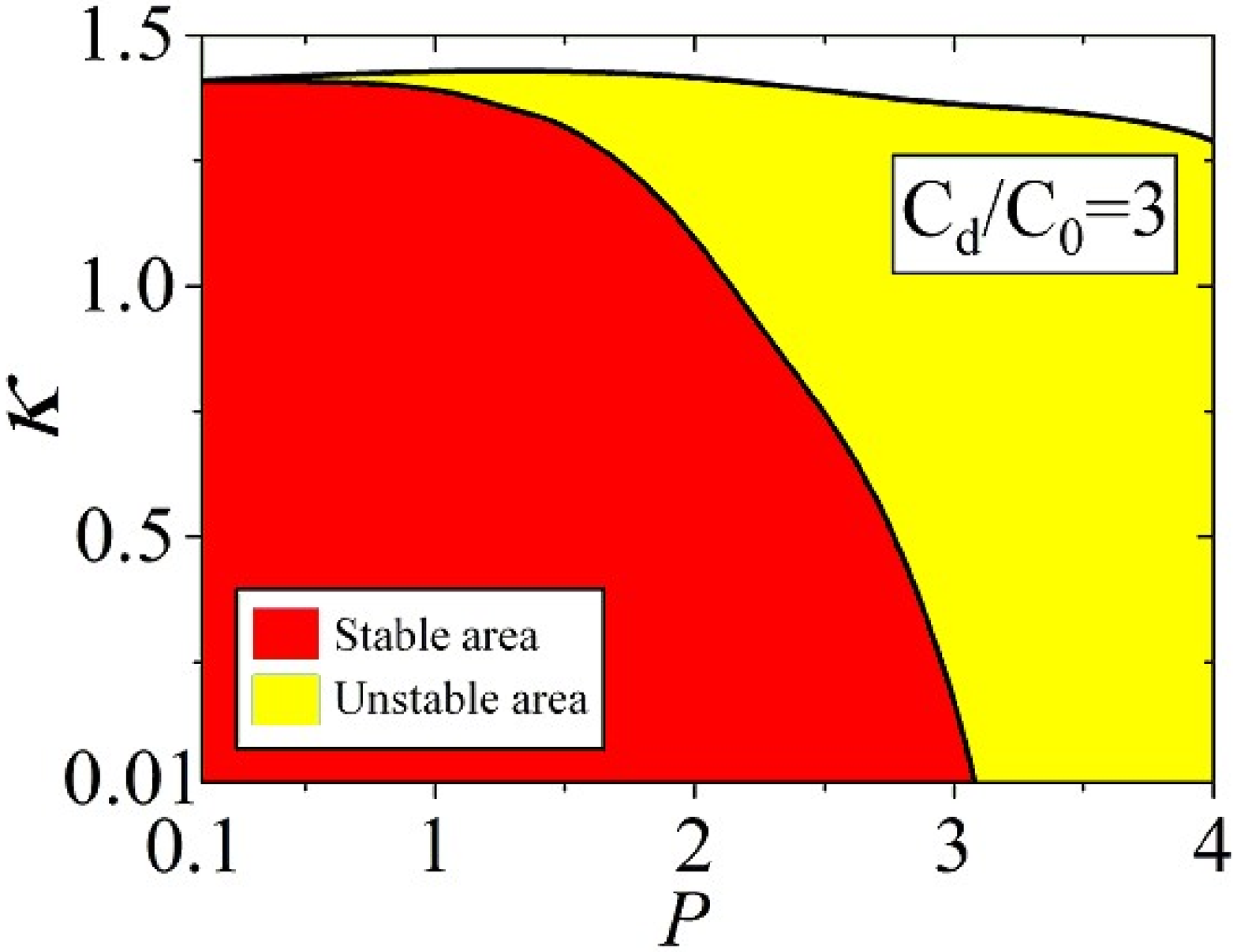}}
\subfigure[]{\label{fig4c}
\includegraphics[scale=0.14]{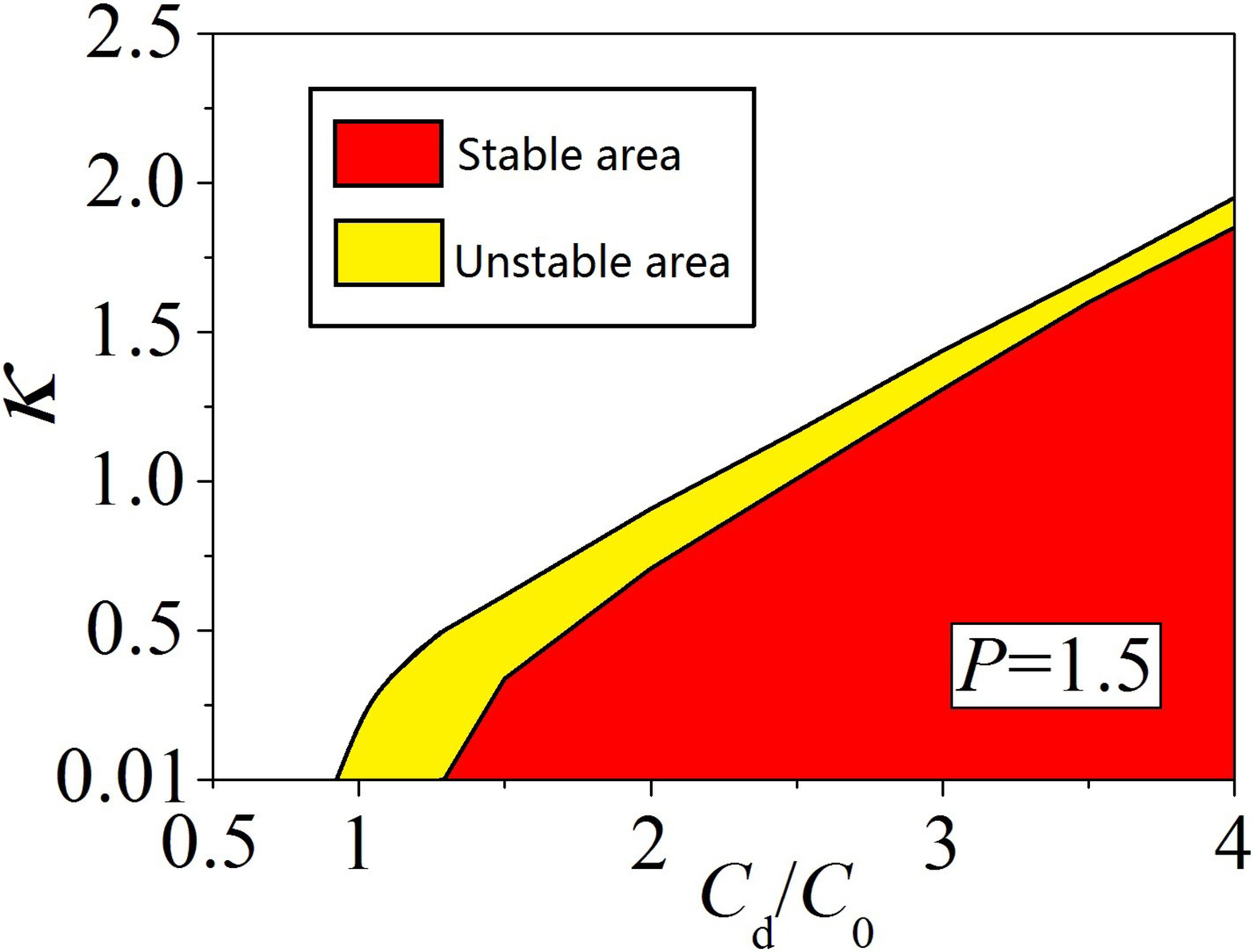}}%
\subfigure[]{\label{fig4c}
\includegraphics[scale=0.28]{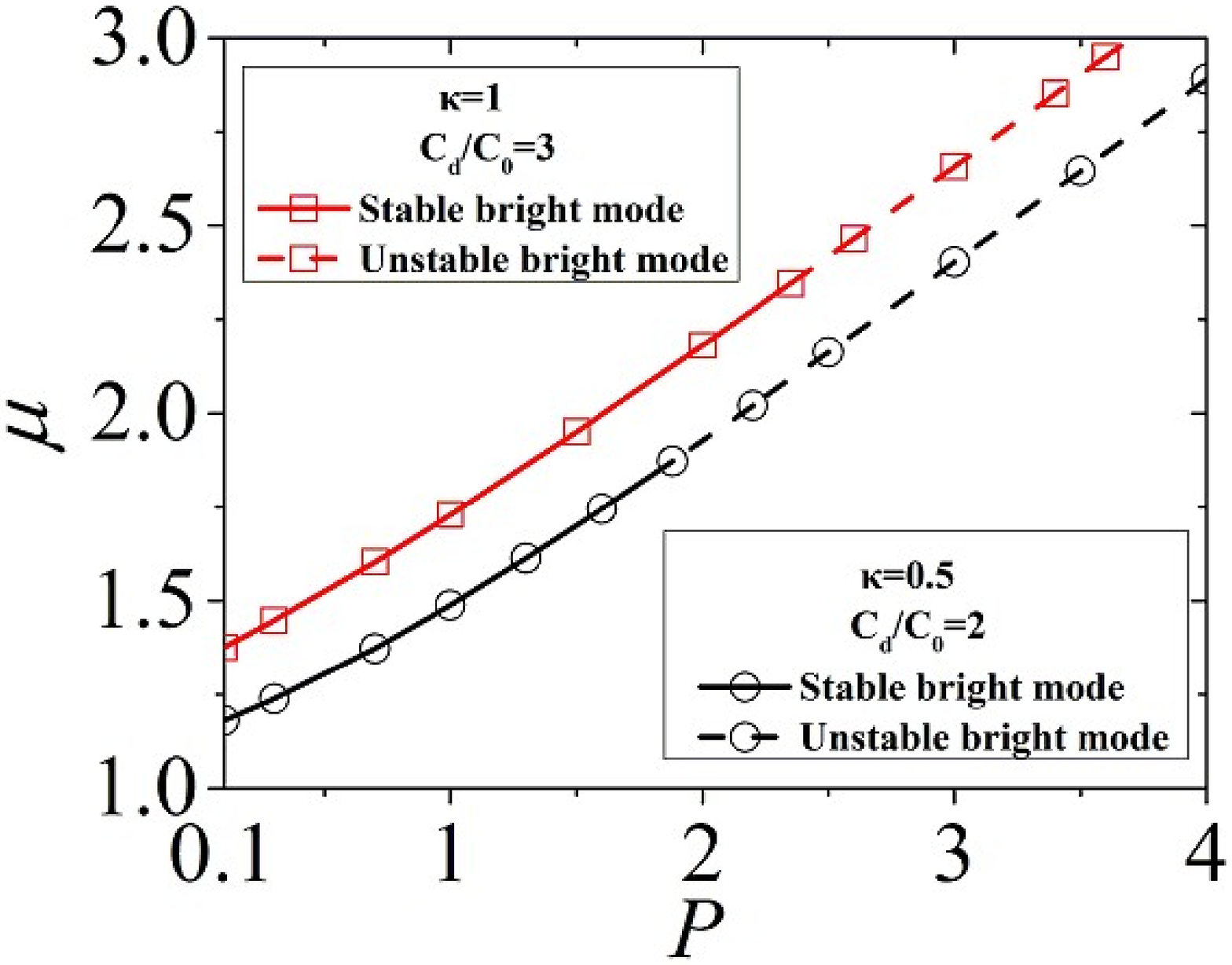}}
\caption{(Color online) Existence regions of stable and unstable bright
staggered solitons in the model based on Eq. (\protect\ref{DNLS}) in the
planes $(\protect\kappa ,P)$ with $C_{d}/C_{0}=2$ (a) and $C_{d}/C_{0}=3$
(b). (c) The stability diagram in the $(\protect\kappa ,C_{d}/C_{0})$ plane
with $P=1.5$. The solitons are stable and unstable, respectively, in red and
yellow areas. No staggered bright solitons have been found in white areas.
(d) Dependence of $\mu(P)$ for the stagger solitons at fixed values of other parameters.}
\label{Distributelinearbight}
\end{figure}

The results for the bright DSs of this type are summarized in stability
charts in parameter planes of $(\kappa ,P)$ and $(\kappa ,C_{d}/C_{0})$,
which are displayed in Fig. \ref{Distributelinearbight} [recall that $P$ is
the total power defined by Eq. (\ref{P})]. The figure demonstrates that the
pinned bright DS gains stability with the increase of the intrinsic coupling
strength of the dimer, $C_{d}$, while the increase of the the gain-loss
coefficient, $\kappa $, naturally leads to destabilization. Indeed, larger
values of $C_{d}$ make the pinning potential (\ref{U}) stronger, and also
facilitate maintaining the balance between the gain and loss, while larger $%
\kappa $ make this harder. It is also seen that there is no minimum
(threshold) value of $P$ necessary for the existence of the staggered bright
DSs.

\subsection{Unstaggered bright modes}

\begin{figure}[tbp]
\centering\subfigure[] {\label{fig5a}
\includegraphics[scale=0.28]{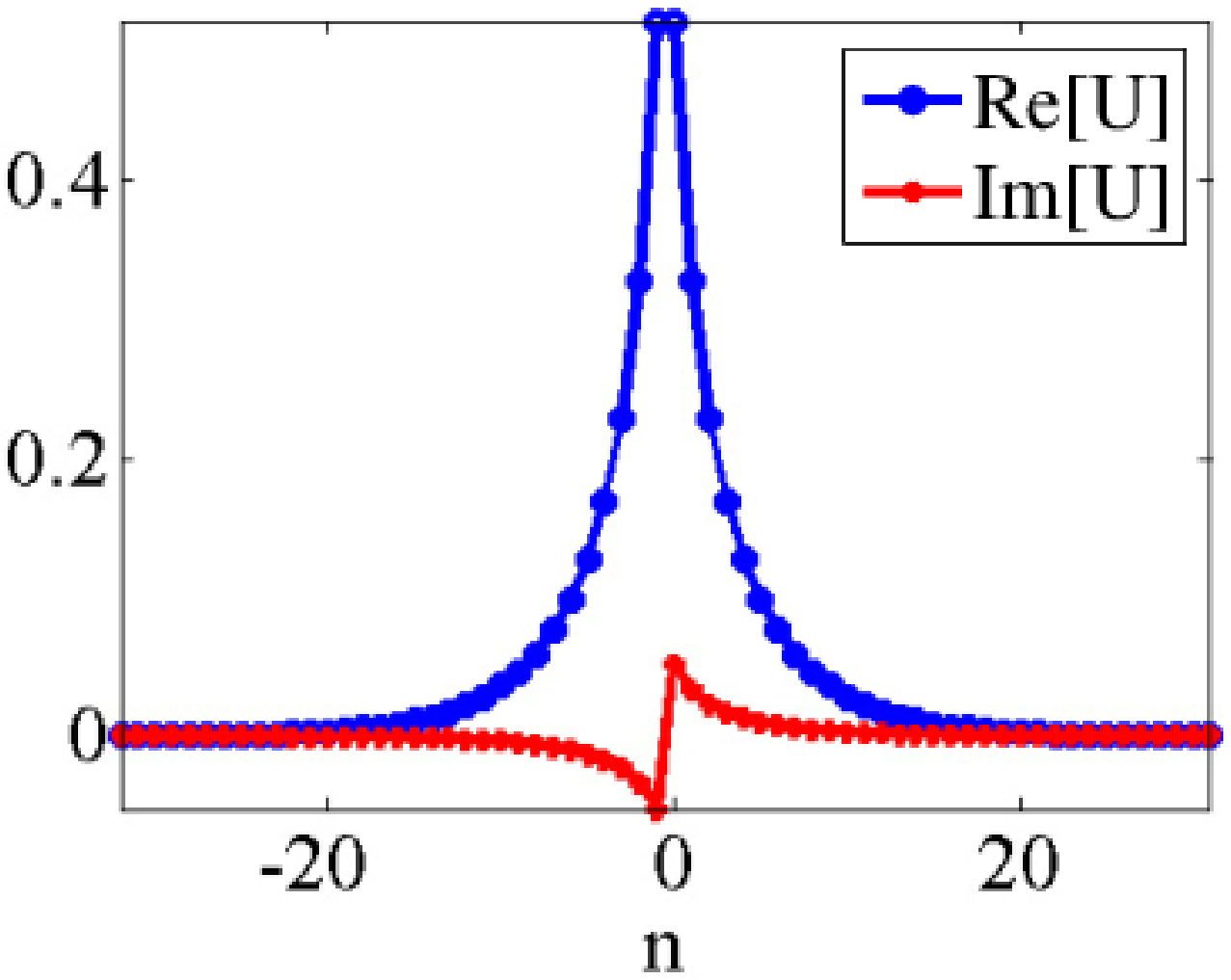}}%
\subfigure[] {\label{fig5b}
\includegraphics[scale=0.28]{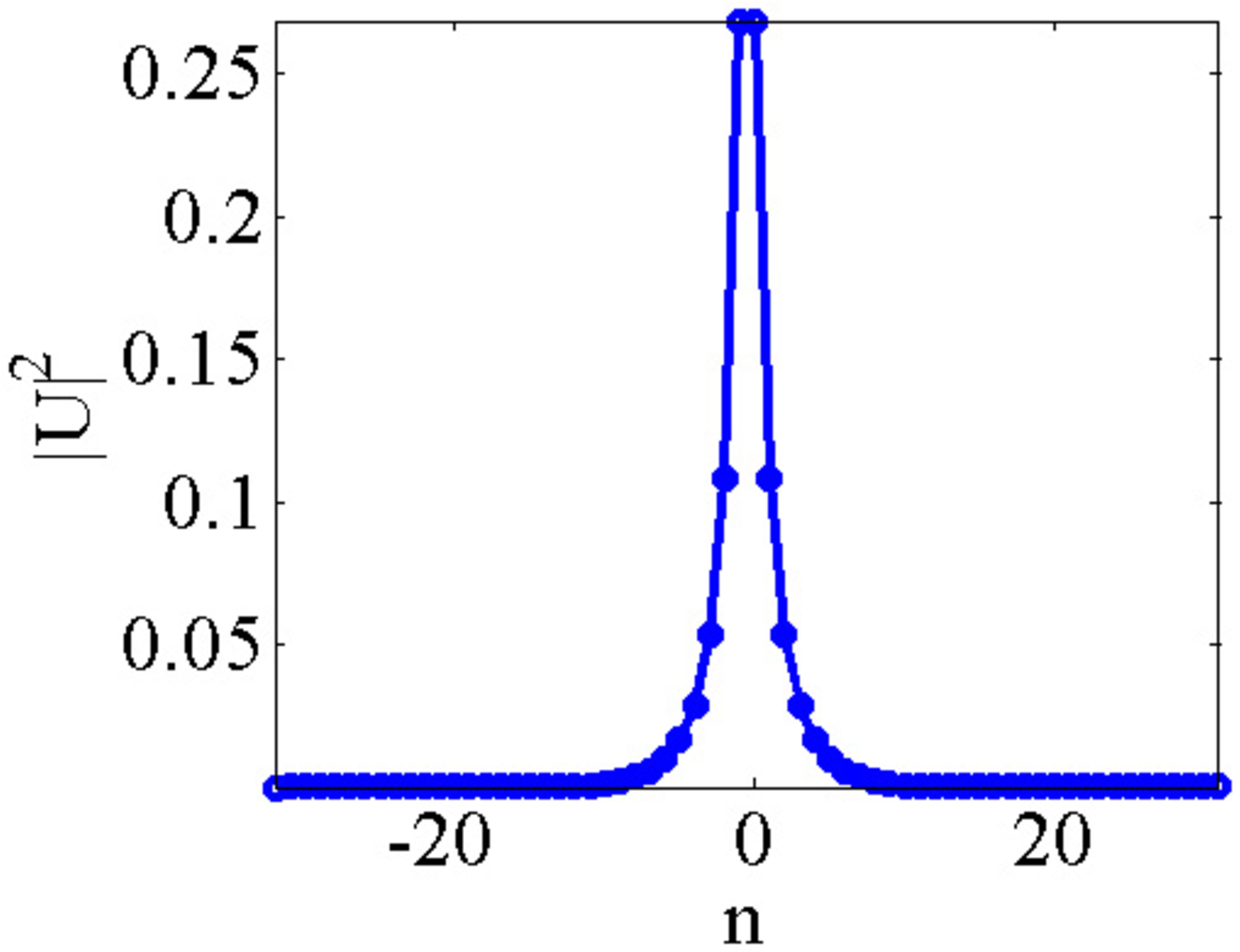}}
\subfigure[]{\label{fig5c}
\includegraphics[scale=0.28]{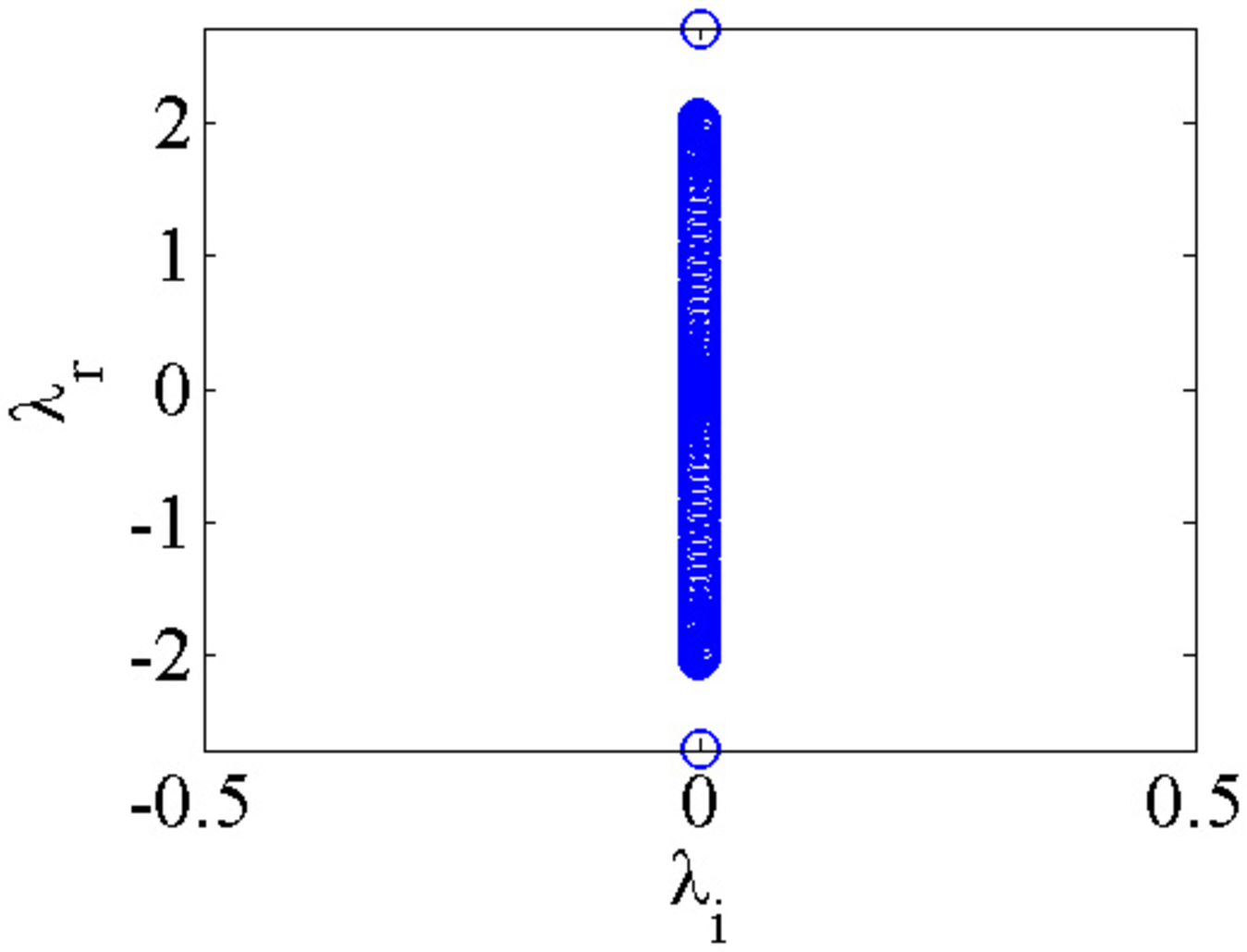}}
\subfigure[]{\label{fig5d}
\includegraphics[scale=0.28]{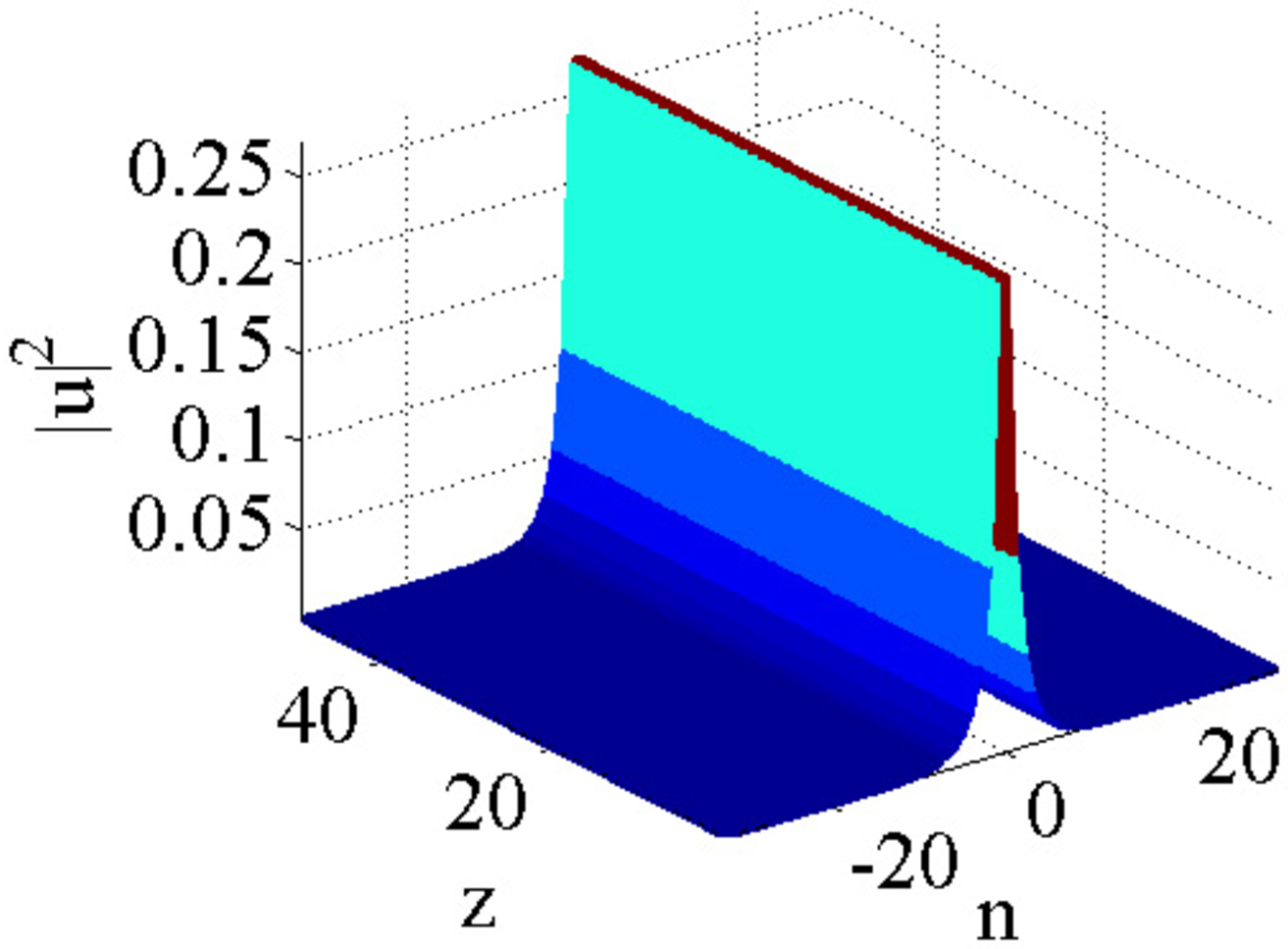}}
\caption{(Color online) A typical example of a bright unstaggered mode
produced by Eq. (\protect\ref{DNLS}) for $\left( P,C_{d}/C_{0},\protect%
\kappa \right) =(1,2,0.1)$. Panels have the same meaning as in Fig. \protect
\ref{stablebright}.}
\label{brightmode}
\end{figure}

\begin{figure}[tbp]
\centering\subfigure[] {\label{fig6a}
\includegraphics[scale=0.14]{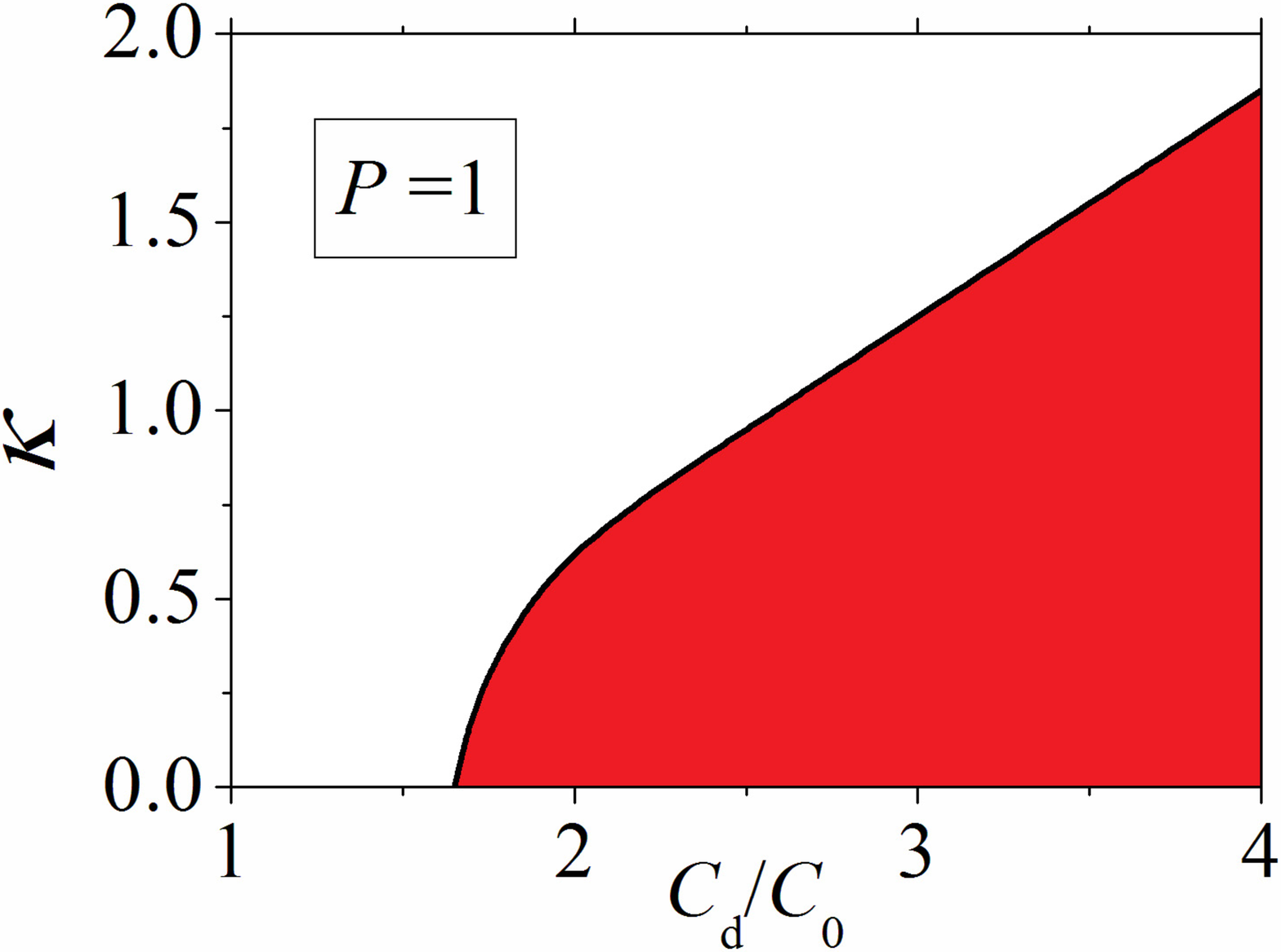}}%
\subfigure[] {\label{fig6b}
\includegraphics[scale=0.21]{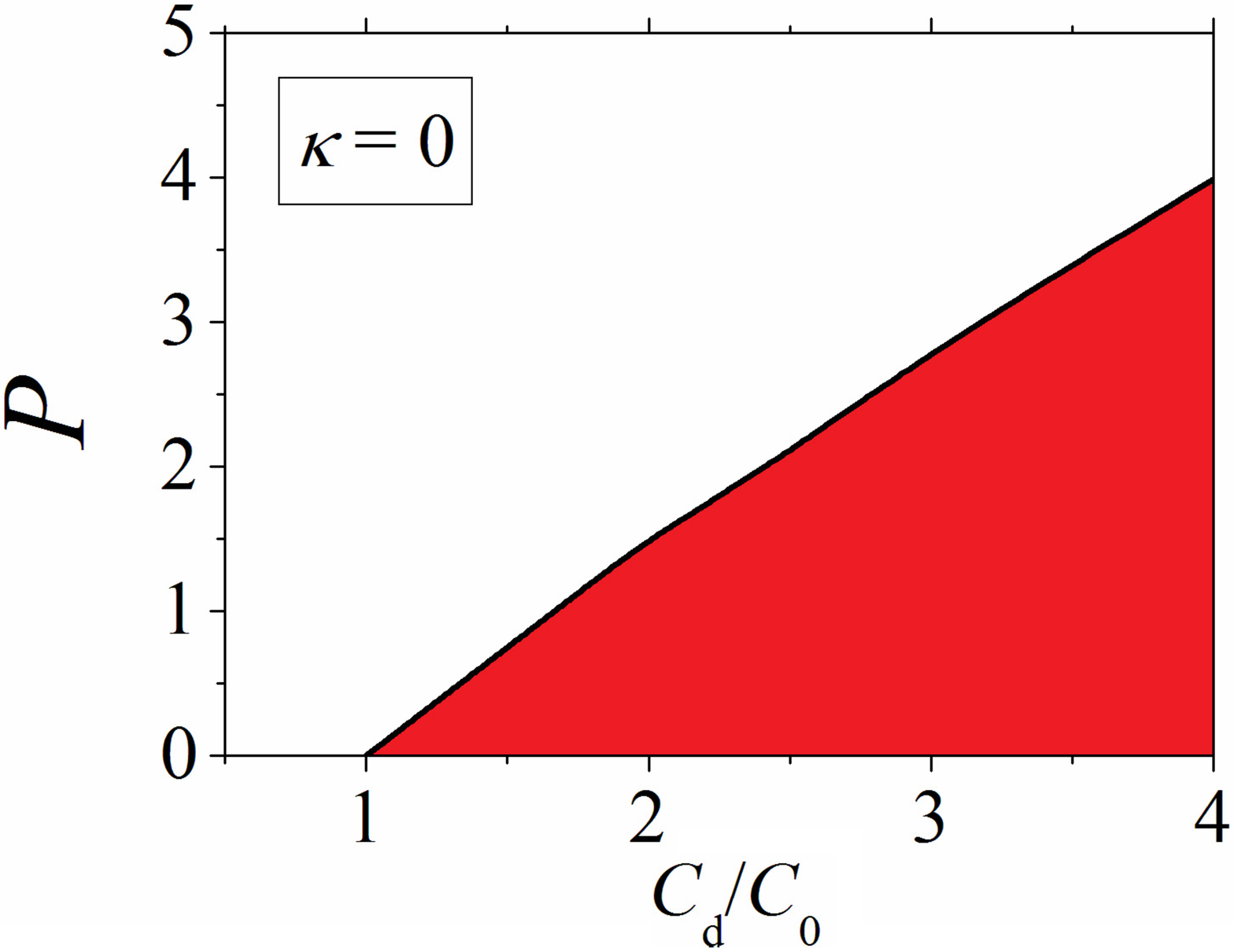}}%
\subfigure[] {\label{fig6b}
\includegraphics[scale=0.29]{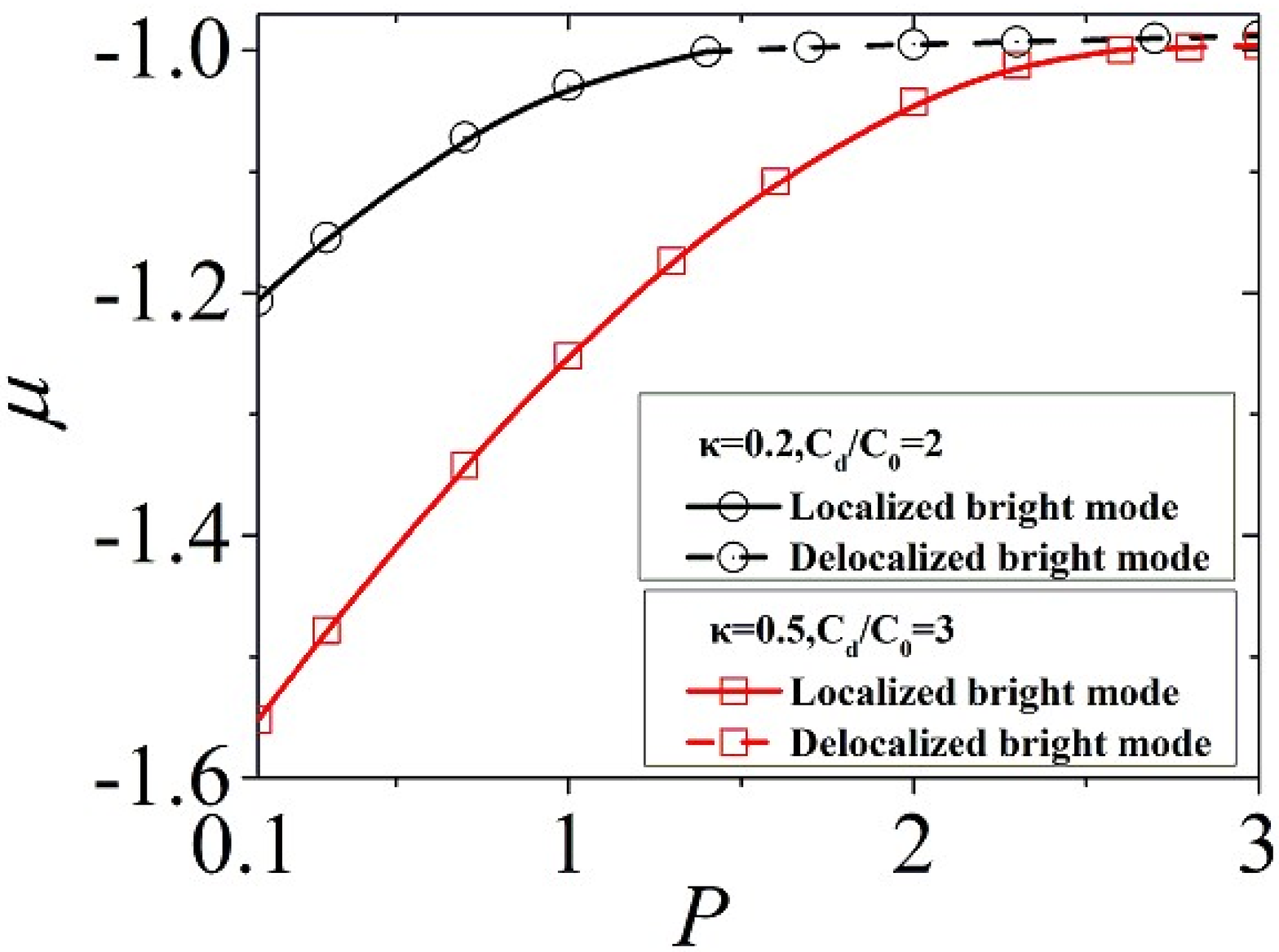}}
\caption{(Color online) The existence area (red) of the unstaggered bright
modes (which are all stable) in the plane of $(\protect\kappa ,C_{d}/C_{0})$ (here
$P=1$ is fixed) (a), and $(P,C_{d}/C_{0})$ (here $\protect\kappa =0$ is fixed)
(b). In the white area, solutions are delocalized. (c) Dependences $\protect%
\mu (P)$ for the modes.}
\label{Brightmodearea}
\end{figure}

Uniform nonlinear waveguide arrays with self-defocusing nonlinearity cannot
support unstaggered bright DSs. However, unstaggered localized modes may
exist, being pinned to the attractive defect. The numerical solution of Eq. (%
\ref{DNLS})\ produces such modes, see an example in Fig. \ref{brightmode}.
They all are \emph{stable}, their existence areas in the planes of $(\kappa
,C_{d}/C_{0})$ and $(P,C_{d}/C_{0})$ being displayed in Fig. \ref{Brightmodearea}.
Similar to the staggered DS, the increase of the intrinsic coupling constant
of the dimer, $C_{d}$, helps to expand the existence area of the bright
modes, which starts from $C_{d}/C_{0}=1$, see Fig. \ref{Brightmodearea}(b).
Note also that, as well as the staggered modes considered above, the
unstaggered ones exhibit no finite existence threshold in terms of the total
power, as seen in Figs. \ref{Brightmodearea}(b,c).

In dependences of the propagation constant on the total power, displayed in
Fig. \ref{Brightmodearea}(c), attaining the level of $d\mu /dP=0$
(designated by horizontal dashed lines) implies a transition to delocalized
states. Actually, these are anti-gray modes considered below.

\section{Gray and anti-gray discrete solitons}

\subsection{Comparison with the continuum-model counterpart}

Gray DSs are solutions to Eq. (\ref{DNLS}) supported by the nonzero
background intensity, $|U_{\mathrm{BG}}|^{2}$, which, in turn, is linked to
the propagation constant by an obvious relation:
\begin{equation}
|U_{\mathrm{BG}}|^{2}=\mu +2C_{0}.  \label{UBG}
\end{equation}%
The interaction of gray solitons with the defect may be estimated, in the
continuum limit, by means of the effective potential (\ref{U}), where a
dark-soliton solution should be substituted. For a typical shape of this
solution, $u_{\mathrm{dark}}=A\tanh \left( a\xi \right) $, Eq. (\ref{U})
yields
\begin{equation}
U_{\mathrm{dark}}(\xi )=\varepsilon A^{2}a^{2}\mathrm{sech}^{4}\left( a\xi
\right) .  \label{max}
\end{equation}%
On the contrary to its counterpart for the bright soliton, given by Eq. (\ref%
{min}), this potential features a maximum at $\xi =0$ for $\varepsilon >0$,
and a minimum for $\varepsilon <0$, hence it may be expected to be
attractive in the latter case, which, as said above, corresponds to $%
C_{d}<C_{0}$. Indeed, at strengths of the gain and loss, $\kappa $, small enough, stable gray solitons pinned to the defect tend to exist at $%
C_{d}/C_{0}<1$, as can be seen below in Fig. \ref{DistributLinear}(b).

As said above, anti-gray solitons feature elevation on top of the finite
background, rather than the dip characteristic to the gray ones. The
estimate based on using the effective potential (\ref{U}) is not relevant
for them, as free anti-gray solitons do not exists in the continuum limit.
In fact, the numerical results presented below reveal their existence, in
the form pinned to the defect in the discrete system, at $C_{d}>C_{0}$ [see
Fig. \ref{DistributLinear}(b)], which would correspond to $\varepsilon >0$
in the continuum limit. On the other hand, it is shown below that the
existence of the anti-gray solitons pinned to the defect can be explained by
means of another (strongly discrete) analytical approximation, see Eqs. (\ref%
{flat})-\ref{boundary}).

\subsection{Numerical results}

To find solutions of the gray and anti-gray types, we used the
imaginary-time-propagation method, fixing the total power as $P=128$, which
is exactly equal to the total number of the lattice sites, $N=128$. If we
neglect a relatively small effect of the soliton's core and boundary
conditions on $P$, the corresponding background level is $|U_{\mathrm{BG}%
}|^{2}\approx P/N=1$, which makes it nearly fixed for the gray and anti-gray
DSs.

\begin{figure}[tbp]
\centering\subfigure[] {\label{fig7a}
\includegraphics[scale=0.27]{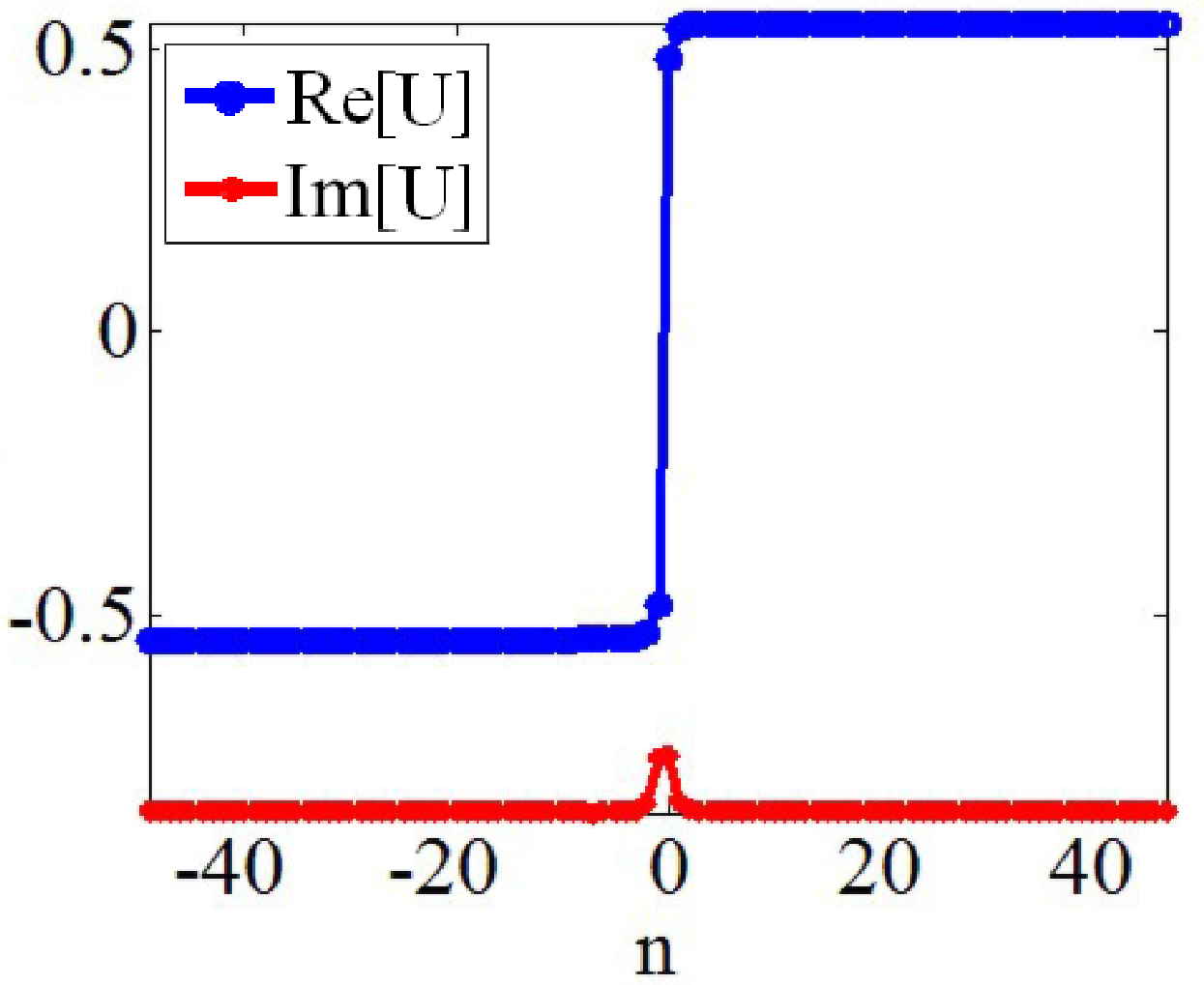}}%
\subfigure[] {\label{fig7b}
\includegraphics[scale=0.2]{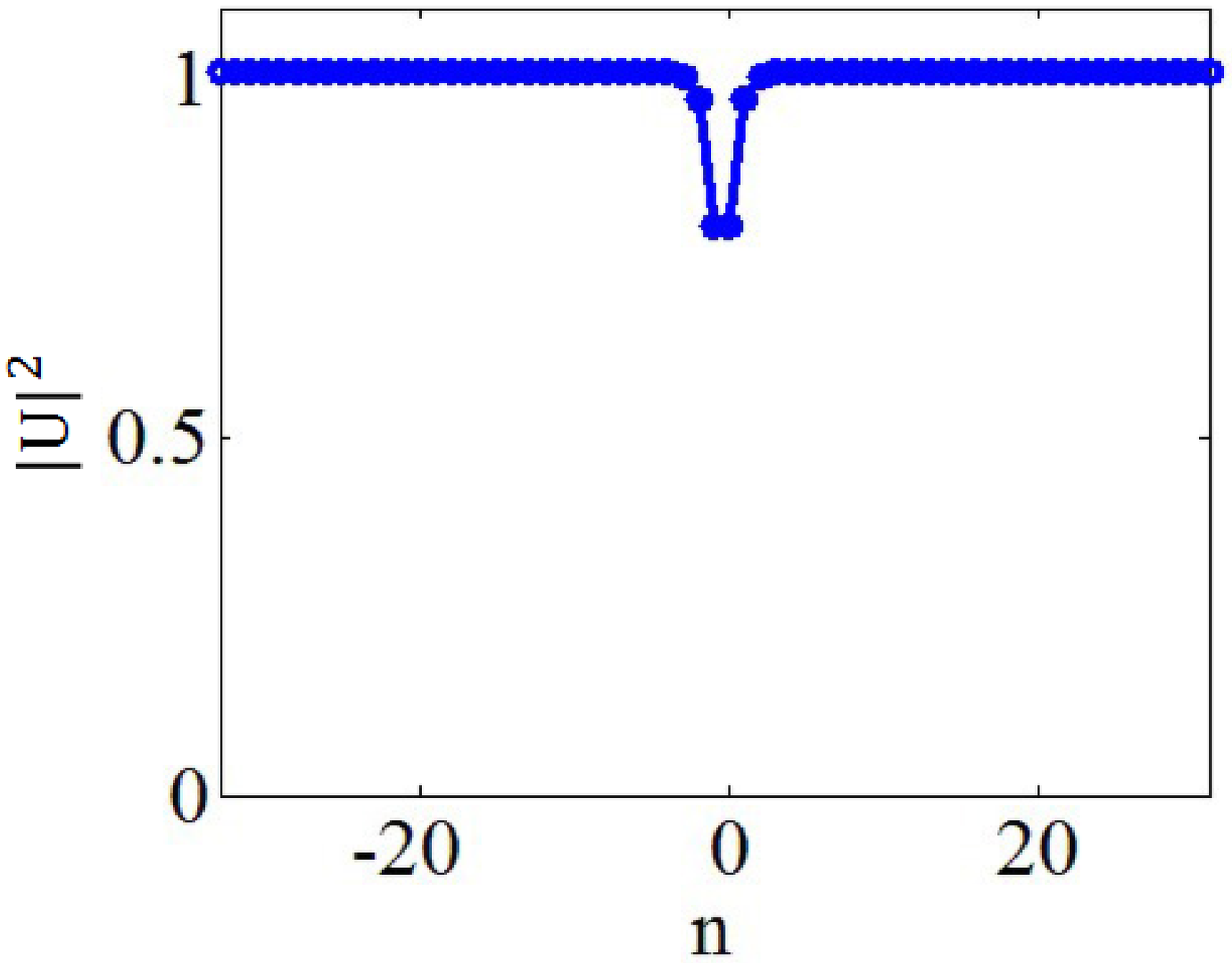}}
\subfigure[]{\label{fig7c}
\includegraphics[scale=0.2]{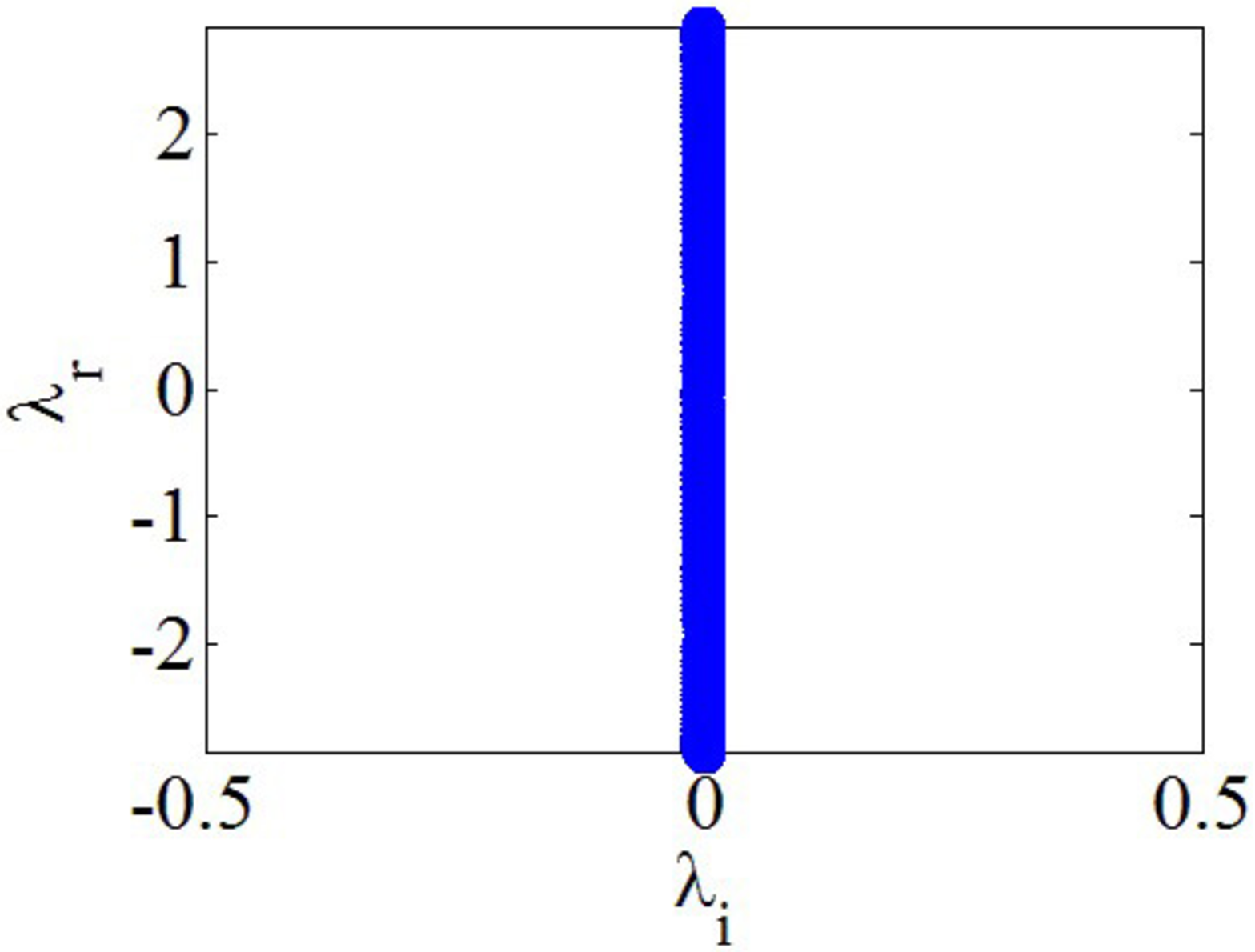}}
\subfigure[]{\label{fig7d}
\includegraphics[scale=0.2]{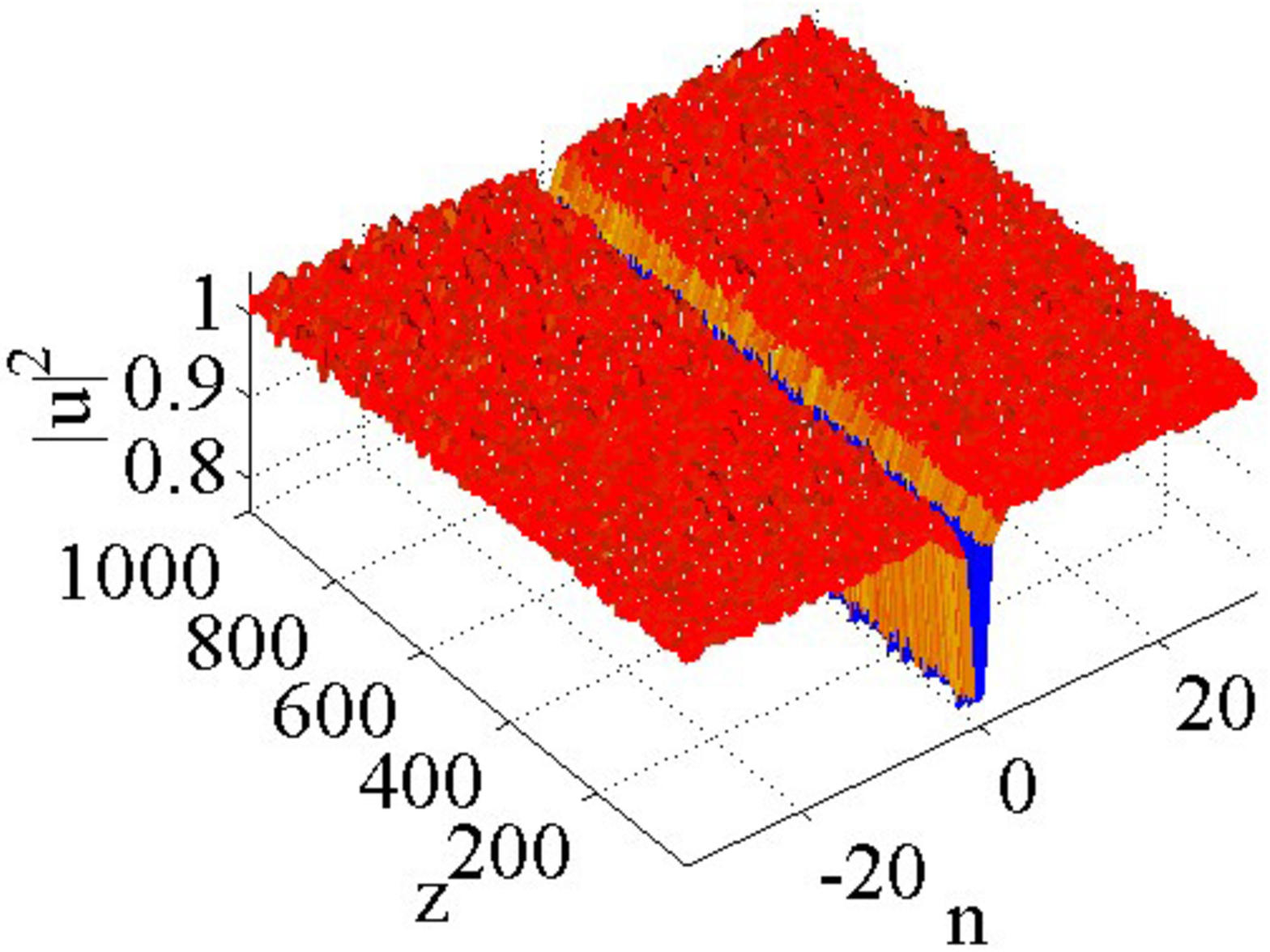}}
\caption{(Color online) A typical example of a stable gray soliton produced
by Eq. (\protect\ref{DNLS}) with $\left( |U_{\mathrm{BG}}|^{2} ,C_{d}/C_{0},\protect%
\kappa \right) =(1,1.1,0.5)$. Panels have the same meaning as in Fig.
\protect\ref{stablebright}.}
\label{LinearGray}
\end{figure}

The numerical analysis has demonstrated that both the gray and anti-gray
DSs, pinned to the $\mathcal{PT}$-dimer defect, are \emph{completely stable}
whenever they exist. The gray DS, supported by the finite background, has a
dip at the center, with a non-zero minimum value, while the anti-gray DS
features a central hump on top of the background. Typical examples of stable
DSs of both types are displayed in Figs. \ref{LinearGray} and \ref%
{LinearAntigray}.
\begin{figure}[tbp]
\centering\subfigure[] {\label{fig8a}
\includegraphics[scale=0.27]{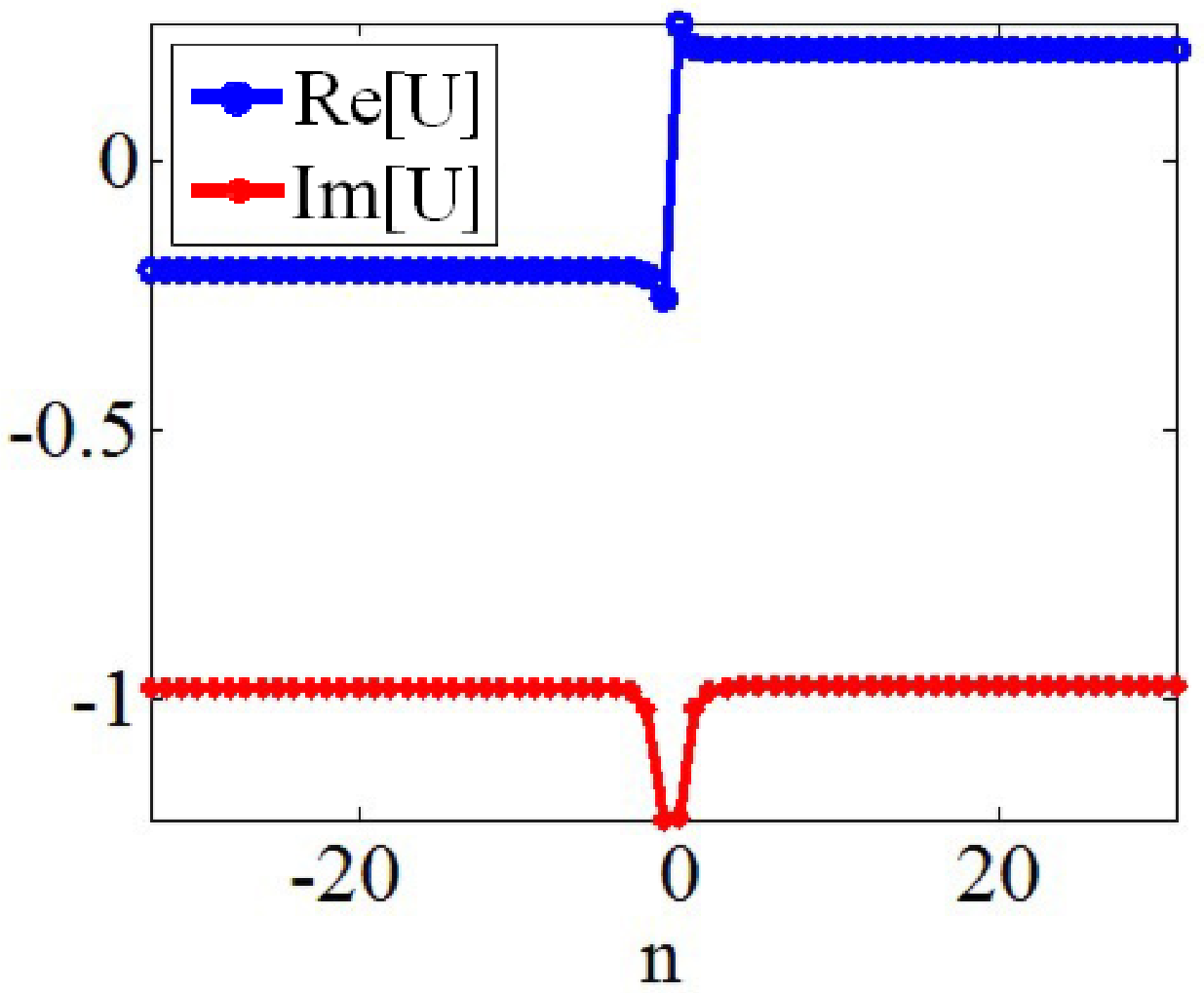}}%
\subfigure[] {\label{fig8b}
\includegraphics[scale=0.2]{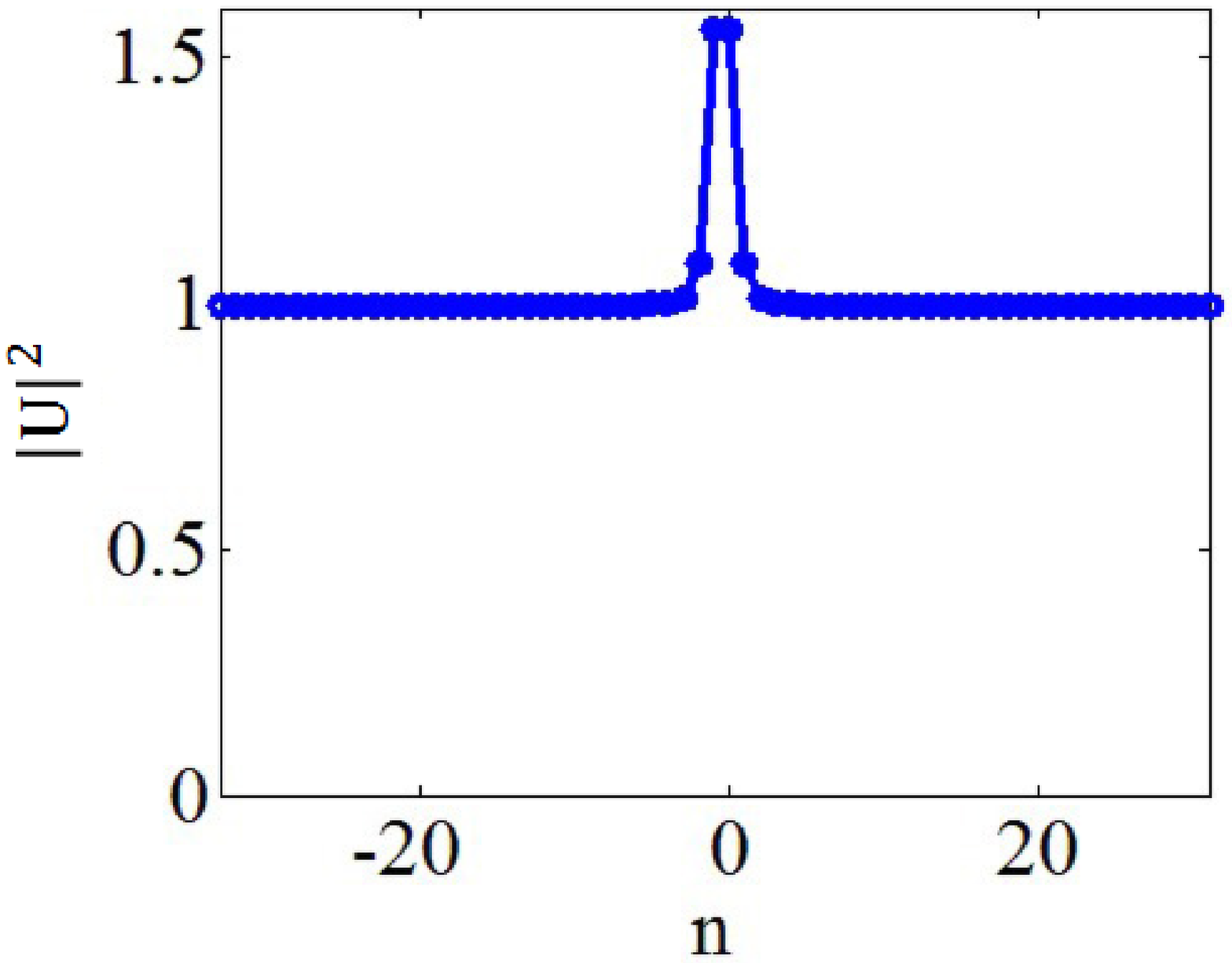}}
\subfigure[]{\label{fig8c}
\includegraphics[scale=0.2]{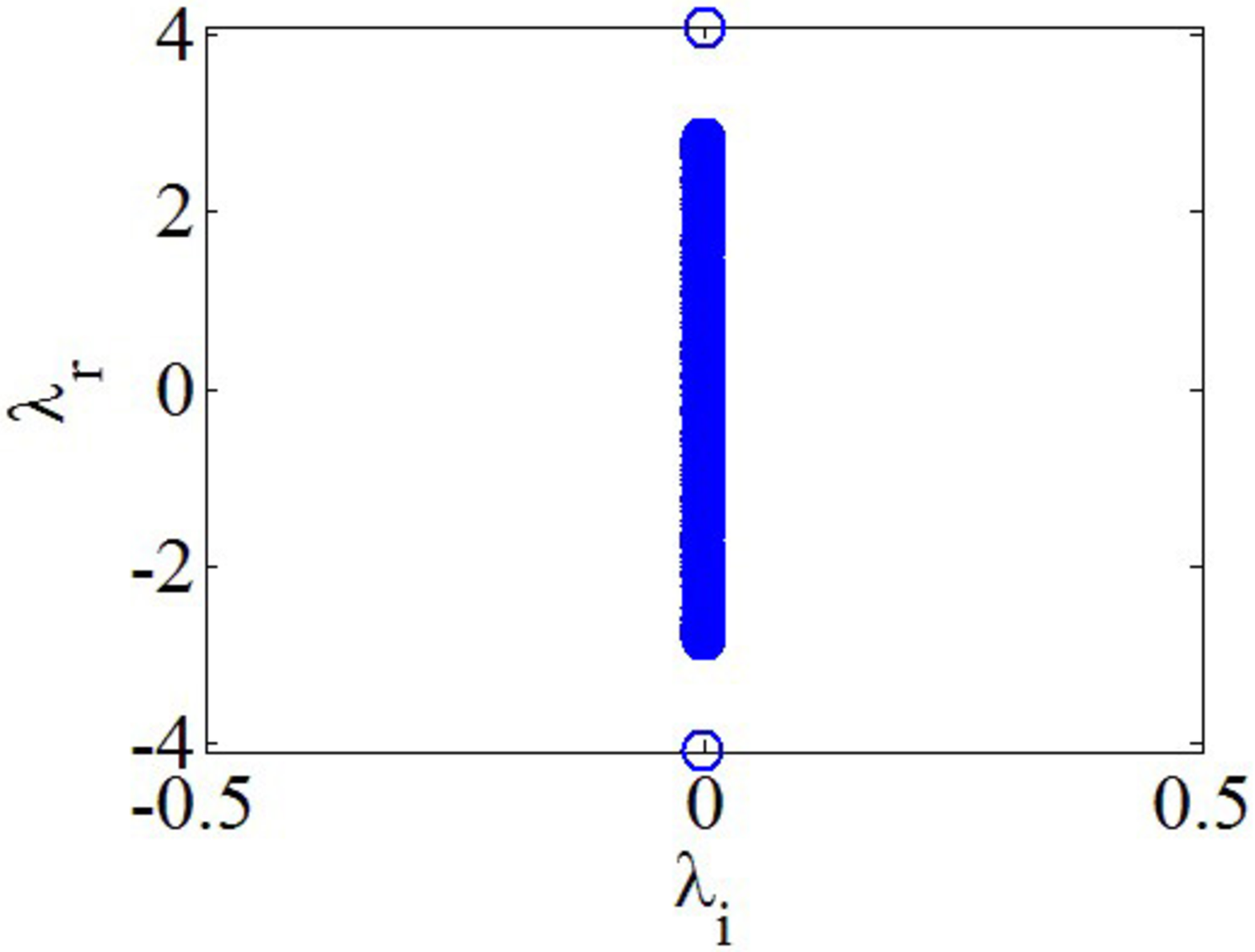}}
\subfigure[]{\label{fig8d}
\includegraphics[scale=0.2]{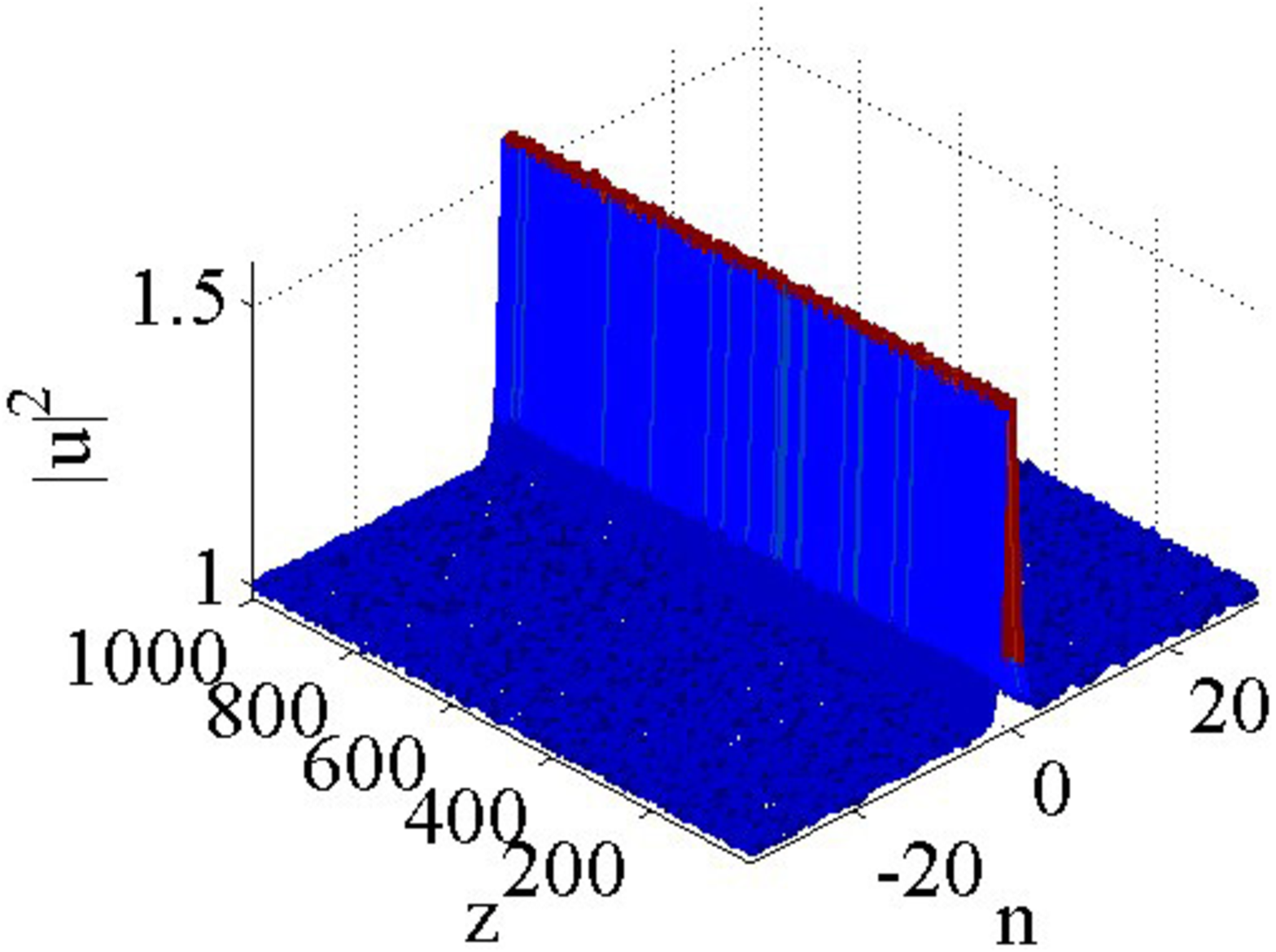}}
\caption{(Color online) A typical example of a stable anti-gray soliton
produced by Eq. (\protect\ref{DNLS}), with $\left( |U_{\mathrm{BG}}|^{2} ,C_{d}/C_{0},%
\protect\kappa \right) =(1,2.5,0.5)$. Panels have the same meaning as in
Fig. \protect\ref{stablebright}.}
\label{LinearAntigray}
\end{figure}

\begin{figure}[tbp]
\centering\subfigure[]{\label{fig9a}
\includegraphics[scale=0.15]{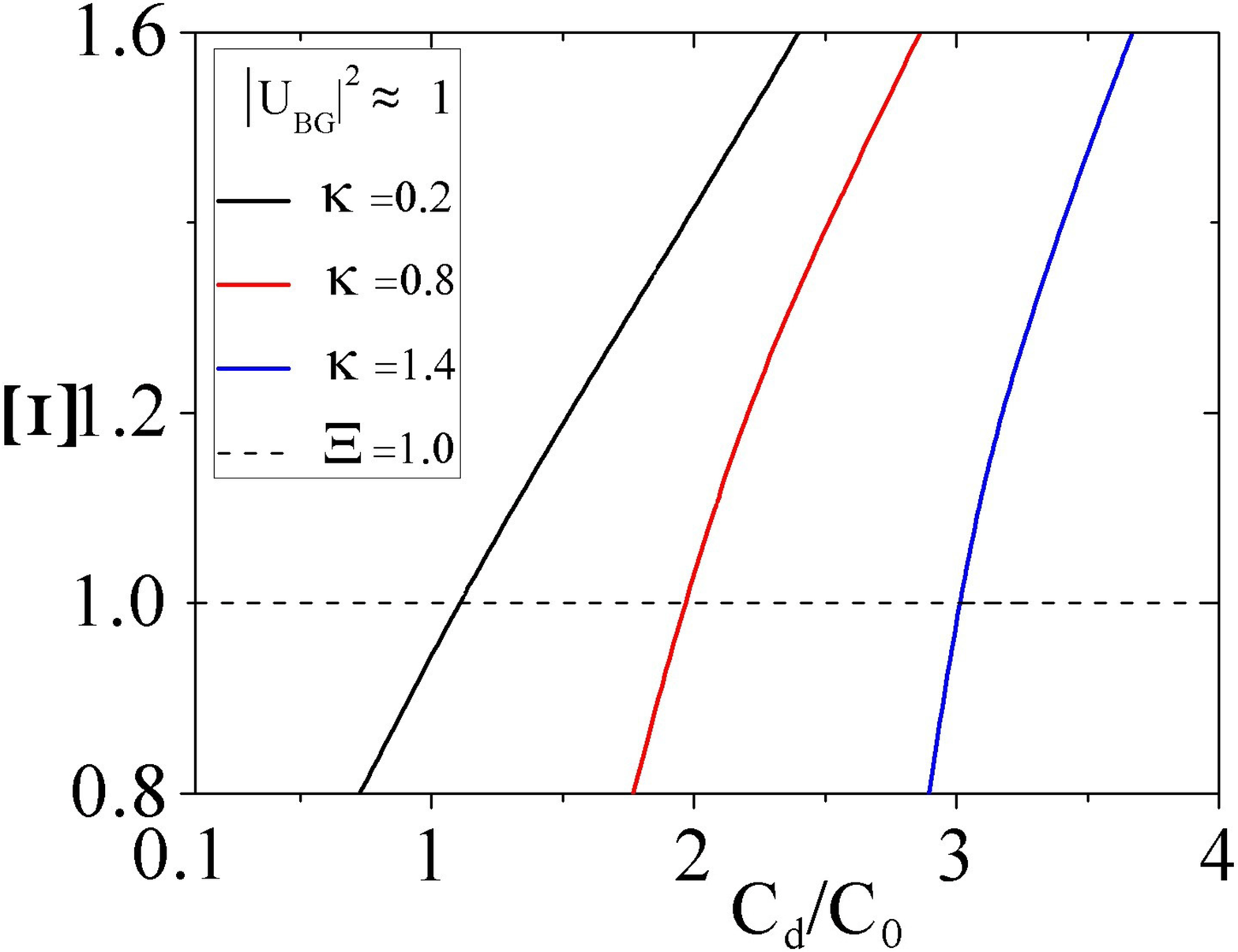}}%
\subfigure[] {\label{fig9b}
\includegraphics[scale=0.22]{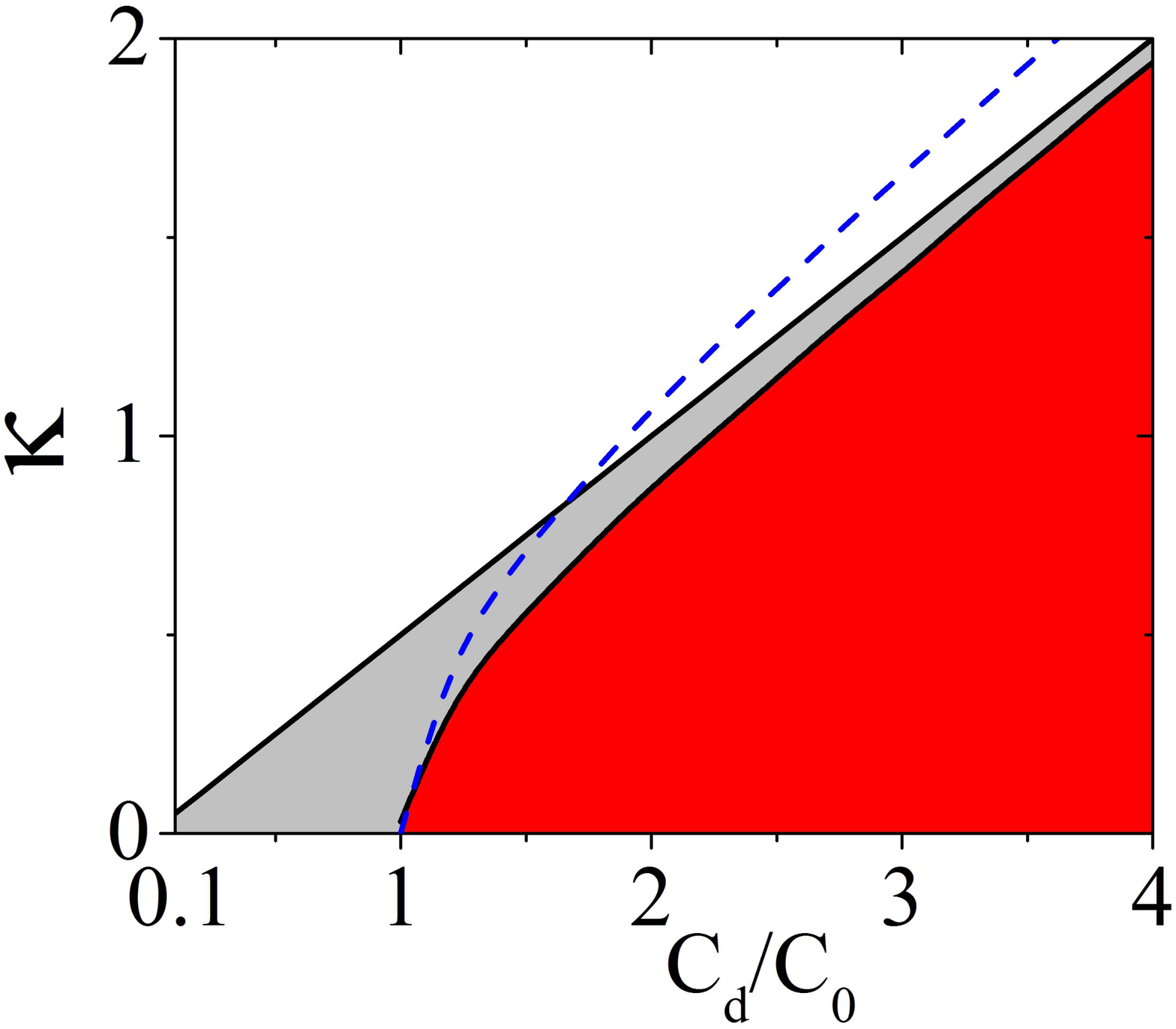}}
\caption{(Color online) (a) The grayness degree, which is defined in Eq. (%
\protect\ref{Gr}), as a function of the dimer's intrinsic coupling constant,
$C_{d}$. The plot comprises both the gray and anti-gray discrete solitons.
The dashed line labels $\Xi =1$. (b) Existence regions of stable gray and
anti-gray discrete solitons (the gray and red areas, respectively) in the $(%
\protect\kappa ,P)$ plane. In the white area, no soliton solutions were
found. Here we fix $P=128$, which corresponds to the background inetnsity $%
|U_{\mathrm{BG}}|^{2}\approx 1$, see the text. The dashed blue curve in (b)
depicts the analytical approximation given by Eq. (\protect\ref{boundary}). }
\label{DistributLinear}
\end{figure}

The solitons of these types are characterized by the ``grayness degree",
\begin{equation}
\Xi ={\frac{|U_{n=-1}|^{2}+|U_{n=0}|^{2}}{2|U_{\mathrm{BG}}|^{2}}}
\label{Gr}
\end{equation}%
where $|U_{\mathrm{BG}}|^{2}$ is the background intensity, given by Eq. (\ref%
{UBG}). Values $\Xi <1$ and $\Xi >1$ imply that the DS is gray or anti-gray,
respectively, while $\Xi \equiv 1$ implies a flat state, which is a border
between them. Figure \ref{fig9a} displays $\Xi $ vs. $C_{d}/C_{0}$ at different
fixed values of $\kappa $.

Stability regions of the gray and anti-gray solitons in the $(\kappa ,C_{d}/C_{0})$
plane are displayed in Fig. \ref{fig9b}. This figure shows that the boundary
between them, $\Xi =1$ [see Eq. (\ref{Gr})], exactly coincides with $%
C_{d}/C_{0}=1$ when $\kappa =0$ (the gain and loss are absent), which is
explained by the fact that the conservative defect is attractive at $%
C_{d}/C_{0}>1$, and repulsive at $C_{d}/C_{0}<1$. The same argument explains
the observation that, at $\kappa >0$, the increase of $C_{d}/C_{0}$ leads to
the expansion of the existence region for the anti-gray DSs, and shrinkage
of that for the gray solitons.

\subsection{An analytical approximation for the discrete system}

The overall existence boundary for the DSs in Fig. \ref{DistributLinear}(b)
is exactly $\kappa =C_{d}$. This feature is explained by the well-known fact
that the $\mathcal{PT}$ symmetry of the dimer is broken at $\kappa >C_{d}$
\cite{Zezyulin,Hadi,KLi}. As shown in Ref. \cite{we}, the same boundary
remains relevant when the dimer is embedded into a linear lattice.

The boundary between the gray and anti-gray DSs in Fig. \ref{DistributLinear}%
(b) can be predicted in an approximate analytical form. Indeed, it follows
from Eq. (\ref{Gr}) that condition $\Xi =1$ implies that $\left\vert
U_{n}\right\vert \equiv |U_{\mathrm{BG}}|,$ which suggests to approximate
the respective solution by ansatz%
\begin{equation}
U_{n}=\left\{
\begin{array}{c}
|U_{\mathrm{BG}}|,~\mathrm{at}~~n<-1~\mathrm{and}~n>0, \\
|U_{\mathrm{BG}}|e^{-i\delta /2},\mathrm{at}~~n=-1, \\
|U_{\mathrm{BG}}|e^{+i\delta /2},\mathrm{at}~~n=0.%
\end{array}%
\right.  \label{flat}
\end{equation}%
The substitution of this ansatz, along with relation (\ref{UBG}), into Eq. (%
\ref{DNLS}), and looking at it solely at the defect-carrying sites, $n=-1$
and $n=0$, leads to an equation for $\delta $ and $\kappa $,%
\begin{equation}
C_{d}e^{-i\delta /2}+C_{0}-2C_{0}e^{i\delta /2}+i\kappa e^{i\delta /2}=0,
\label{complex}
\end{equation}%
the solution of which is%
\begin{equation}
\delta =2\arctan \left( \frac{\kappa }{2C_{0}+C_{d}}\right) ,  \label{delta}
\end{equation}%
\begin{equation}
\kappa ^{2}=\frac{1}{2}\left[ \sqrt{%
C_{0}^{4}+8C_{0}^{2}C_{d}^{2}+16C_{0}^{3}C_{d}}-\left(
7C_{0}^{2}-2C_{d}^{2}\right) \right] .  \label{boundary}
\end{equation}%
In the limit of $C_{d}\rightarrow \infty $, Eq. (\ref{boundary}) simplifies
to $\kappa \approx C_{d}$. The blue dashed curve in Fig. \ref{fig9b}
displays relation (\ref{boundary}), demonstrating that it produces a
reasonable, although not very accurate, approximation\textbf{.}

%Another existence limit, the border between the white and gray color area in
%Fig. \ref{fig9b}, can be found analytically in the limit of $|U_{\mathrm{BG}%
%}^{2}|\rightarrow 0$, when Eq. (\ref{DNLS}) becomes linear. As shown in Ref.
%\cite{we}, in this case an exact solution can be found as,%
%\begin{equation}
%U_{n}=\left\{
%\begin{array}{c}
%A\exp \left( -i\delta _{0}/2+\lambda (n-N/2)\right) ~\mathrm{at}~~n\leq N/2,
%\\
%A\exp \left( +i\delta _{0}/2+\lambda (N/2+1-n)\right) ~\mathrm{at}~~n\geq
%N/2,%
%\end{array}%
%\right.  \label{linear-mode}
%\end{equation}%
%where $A$ is an arbitrary amplitude, and the phase shift $\delta _{0}$ and
%spatial-decay rate $\lambda $ are determined by the complex equation, $%
%C_{d}e^{+i\delta _{0}/2}-C_{0}e^{\lambda -i\delta _{0}/2}=i\kappa $, i.e.,%
%\begin{equation}
%\lambda =\ln \left( \sqrt{C_{d}^{2}-\kappa ^{2}}/C_{0}\right) ,~\delta
%_{0}=\arcsin \left( \kappa /C_{d}\right) .  \label{linear}
%\end{equation}%
%The condition that the argument of $\arcsin $ in Eq. (\ref{linear}) must not
%exceed $1$ gives rise to boundary $\kappa \leq C_{d}$, which explains the
%diagonal line bordering the existence area in Fig. \ref{fig9b}.

\subsection{Gray and anti-gray discrete solitons in the system with the
nonlinear $\mathcal{PT}$ symmetry}

For bright DS modes, both staggered and unstaggered ones, the consideration
of the model based on Eq. (\ref{DNLS_N}) with the defect carrying the
nonlinear $\mathcal{PT}$ symmetry (NPTS) produces results which are not
qualitatively different from those reported above for its linear-$\mathcal{PT%
}$-symmetry counterpart, therefore we do not discuss them in detail here.
However, new features appear in the NPTS system for gray and anti-gray DSs:
while, as shown above, they are completely stable in the case of the dimer
with the linear $\mathcal{PT}$ symmetry, the NPTS version generates a
nontrivial boundary in the parameter space between stable and unstable
solitons of these types. These results are summarized in Fig. \ref%
{DistributnonLinear}.
\begin{figure}[tbp]
\centering\includegraphics[scale=0.4]{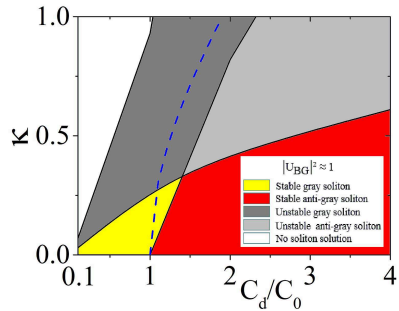}
\caption{(Color online) Stability regions (yellow and red) of gray and
anti-gray solitons in the $(\protect\kappa ,C_{d}/C_{0})$ plane, at $|U_{\mathrm{BG%
}}|^{2}\approx 1$ (the approximate equality is understood here in the same
sense as before, i.e., $P=128$ for the system built of $N=128$ sites). In
dark and light gray areas, respectively, unstable gray and anti-gray
solitons have been found. In the white area, no soliton solutions exist. The
dashed blue curve depicts the analytical approximation (\protect\ref%
{boundary}) for the boundary between gray and anti-gray solitons.}
\label{DistributnonLinear}
\end{figure}

The analytical result for the model with the linear $\mathcal{PT}$-symmetric
dimer, represented by Eq. (\ref{boundary}),
can be easily generalized for the NPTS system, replacing $\kappa $ in those
results by $\kappa |U_{\mathrm{BG}}|^{2}$, pursuant to Eqs. (\ref{DNLS_N})
and (\ref{kappa}). In particular, for the latter system with $|U_{\mathrm{BG}%
}|^{2}=1$, which is represented by Fig. \ref{DistributnonLinear}, the
boundary between the areas of gray and anti-gray DSs is approximated by the
same equation (\ref{boundary}) as above, which is shown by the blue dashed
curve in Fig. \ref{DistributnonLinear}.

\begin{figure}[tbp]
\centering\subfigure[] {\label{NPT8a}
\includegraphics[scale=0.27]{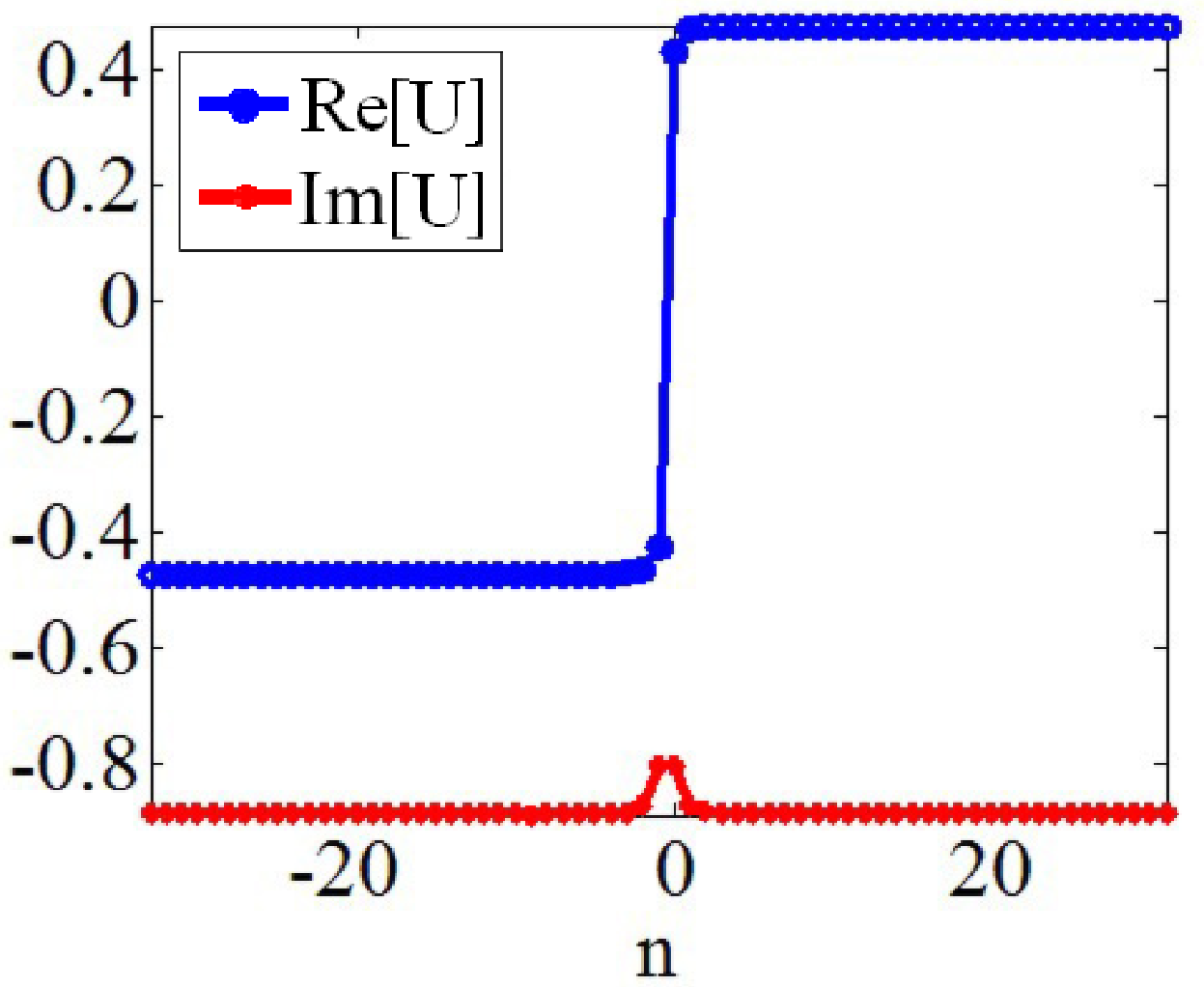}}%
\subfigure[] {\label{NPT8b}
\includegraphics[scale=0.2]{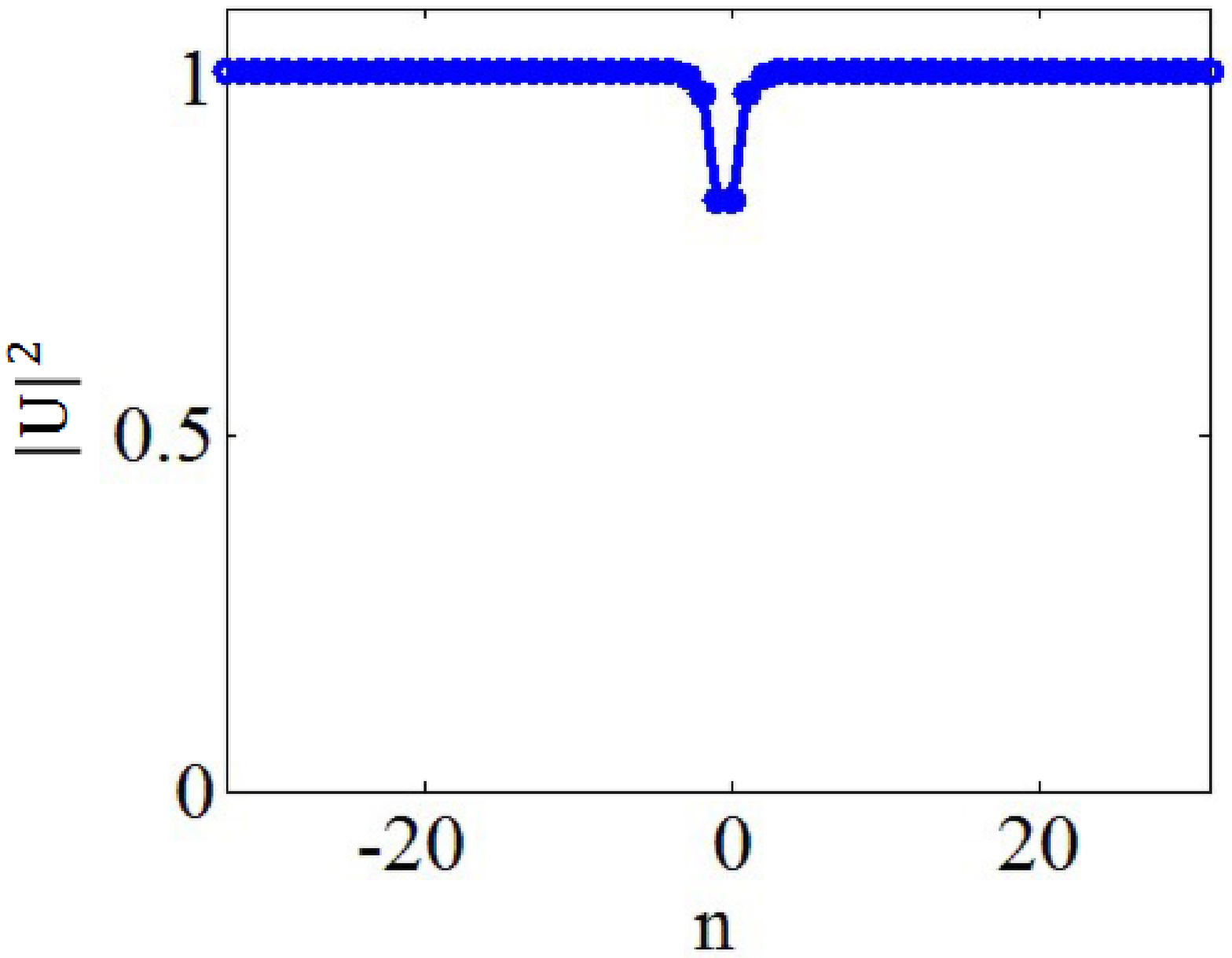}}
\subfigure[]{\label{NPT8c}
\includegraphics[scale=0.2]{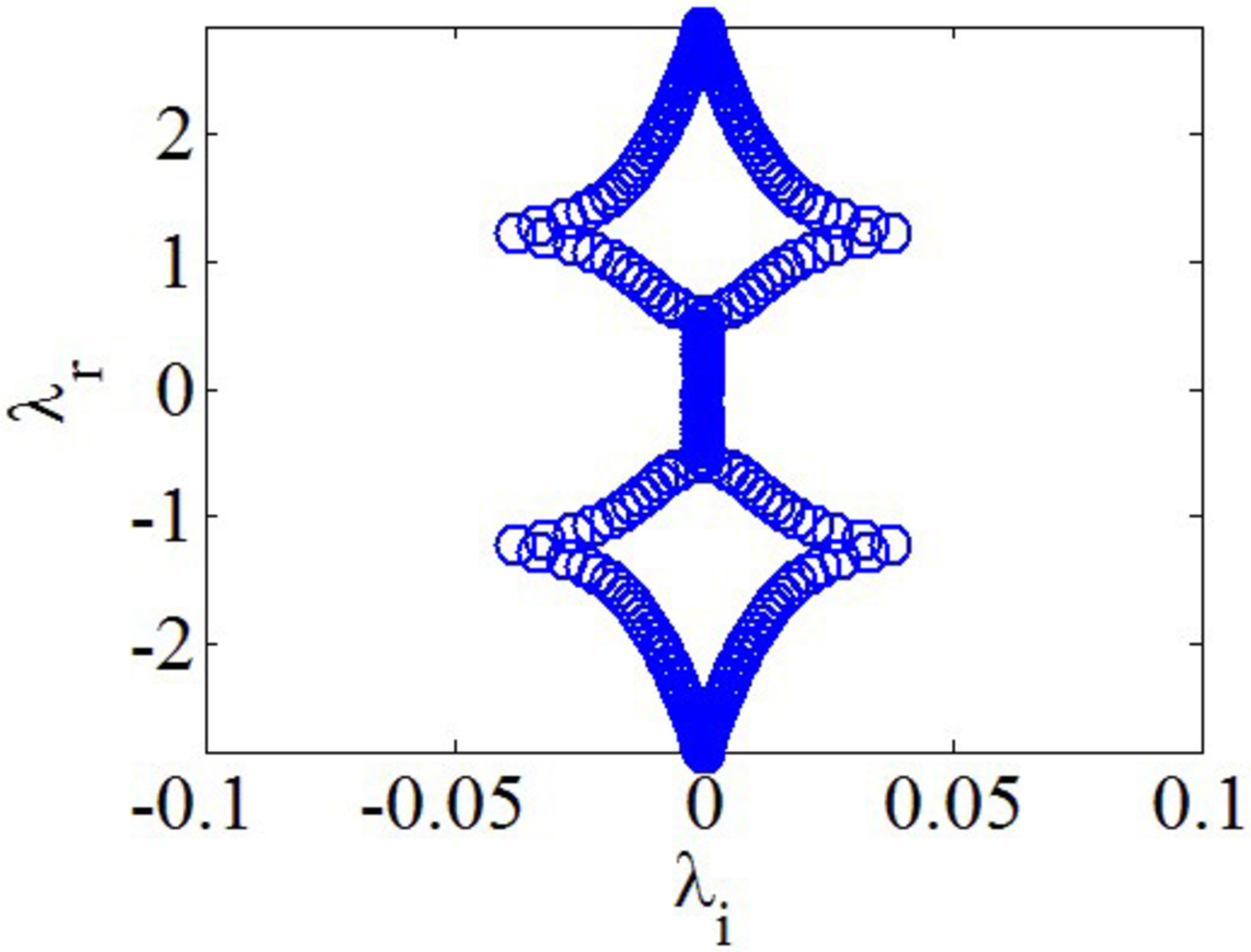}}
\subfigure[]{\label{NPT8d}
\includegraphics[scale=0.2]{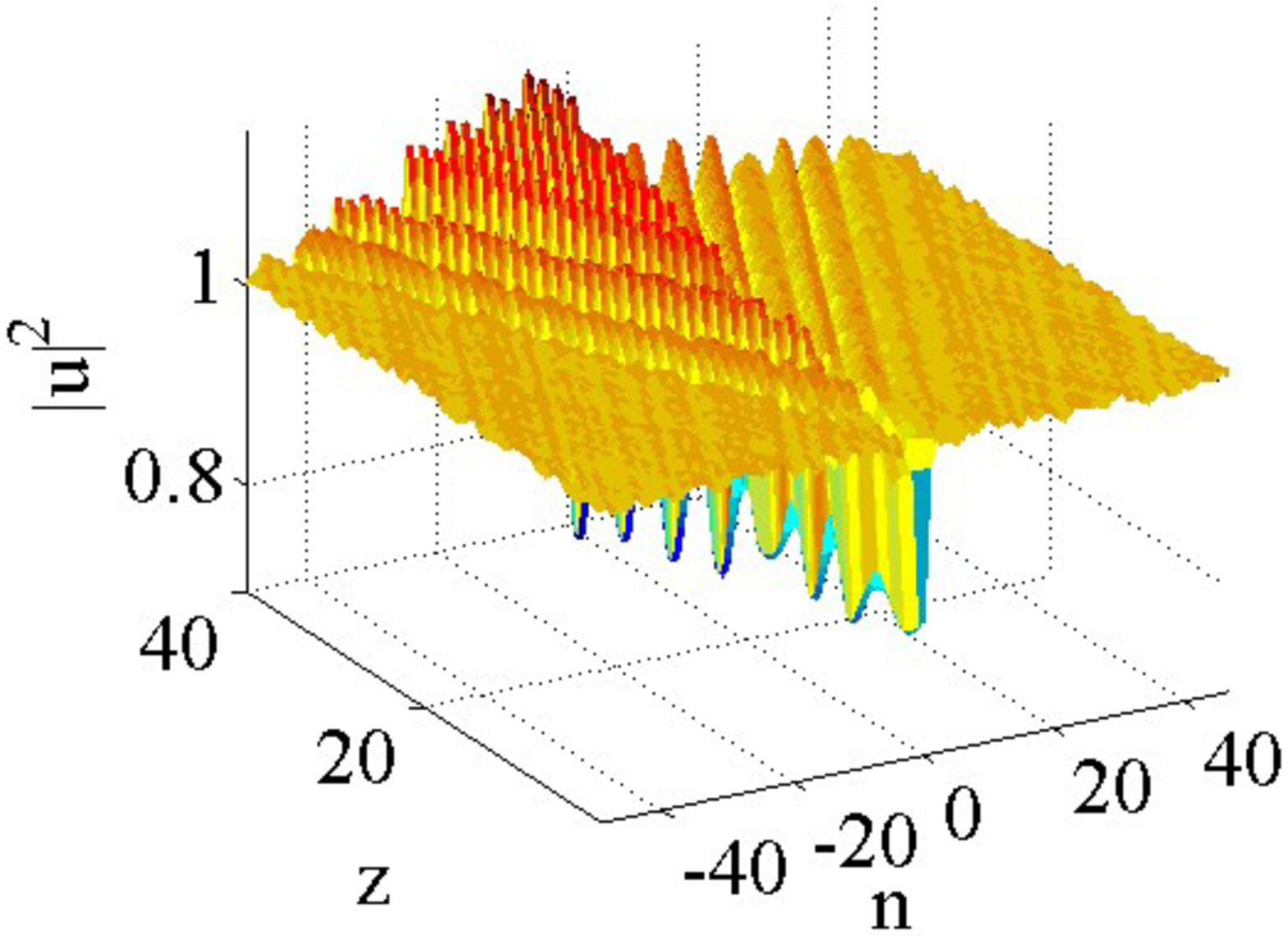}}
\caption{(Color online) A typical example of an unstable gray soliton with $%
\left( |U_{\mathrm{BG}}|^{2},C_{d}/C_{0},\protect\kappa \right) =(1,1,0.5)$
in the model with the nonlinear $\mathcal{PT}$ symmetry of the embedded
dimer, based on Eq. (\protect\ref{DNLS_N}). The panels have the same meaning
as in Fig. \protect\ref{stablebright}.}
\label{unstablenonLinearGray}
\end{figure}

\begin{figure}[tbp]
\centering\subfigure[] {\label{NPT9a}
\includegraphics[scale=0.27]{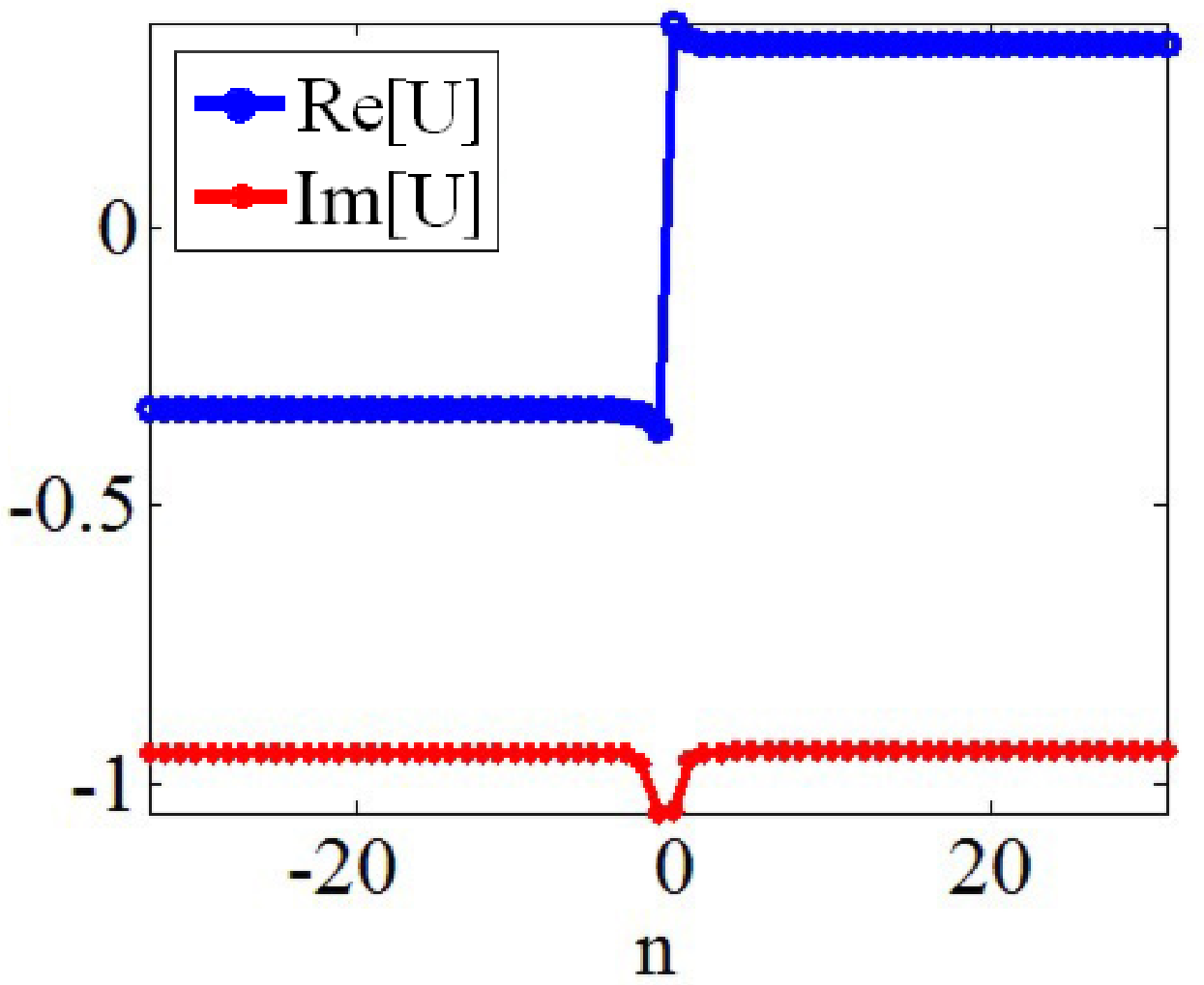}}%
\subfigure[] {\label{NPT9b}
\includegraphics[scale=0.2]{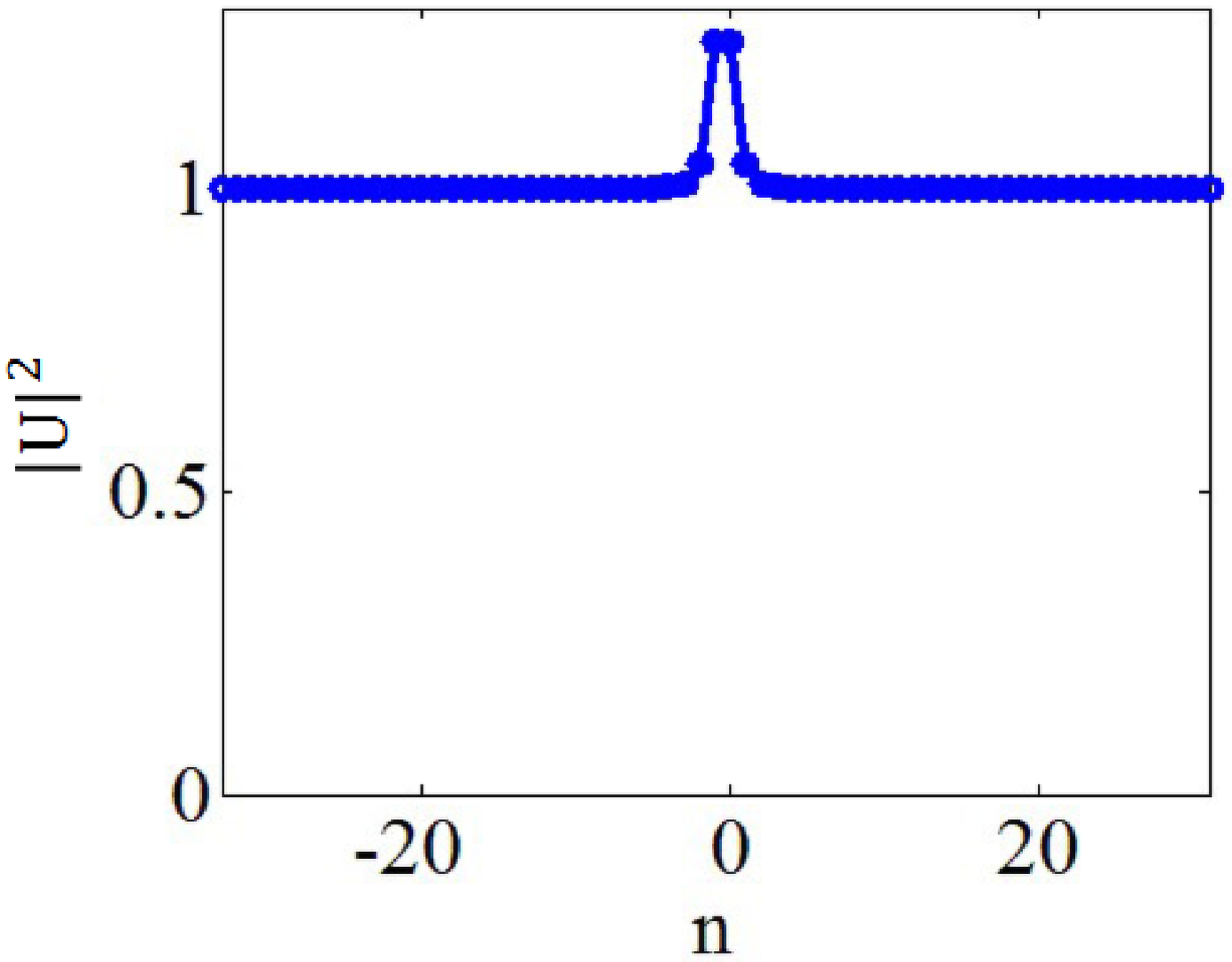}}
\subfigure[]{\label{NPT9c}
\includegraphics[scale=0.2]{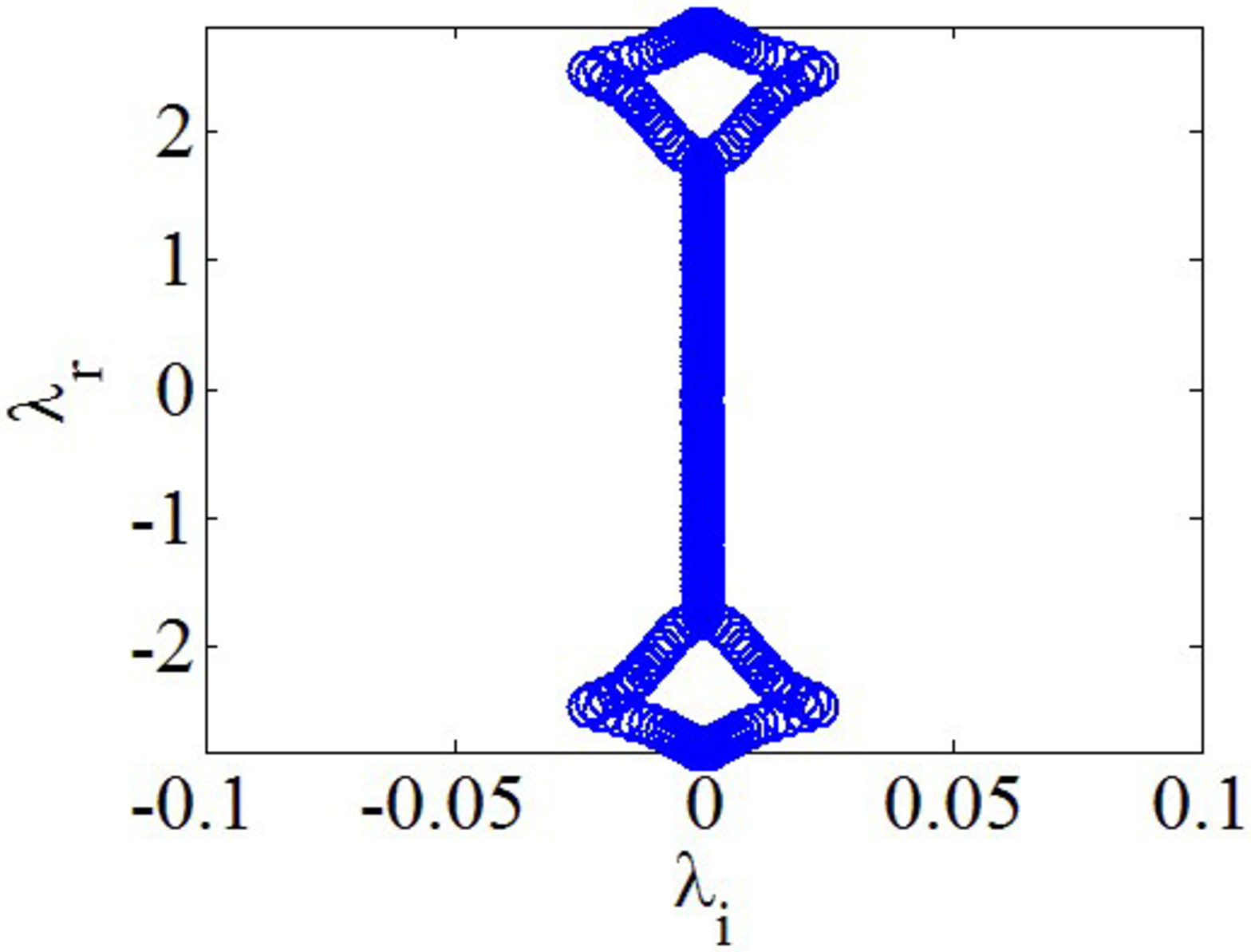}}
\subfigure[]{\label{NPT9d}
\includegraphics[scale=0.2]{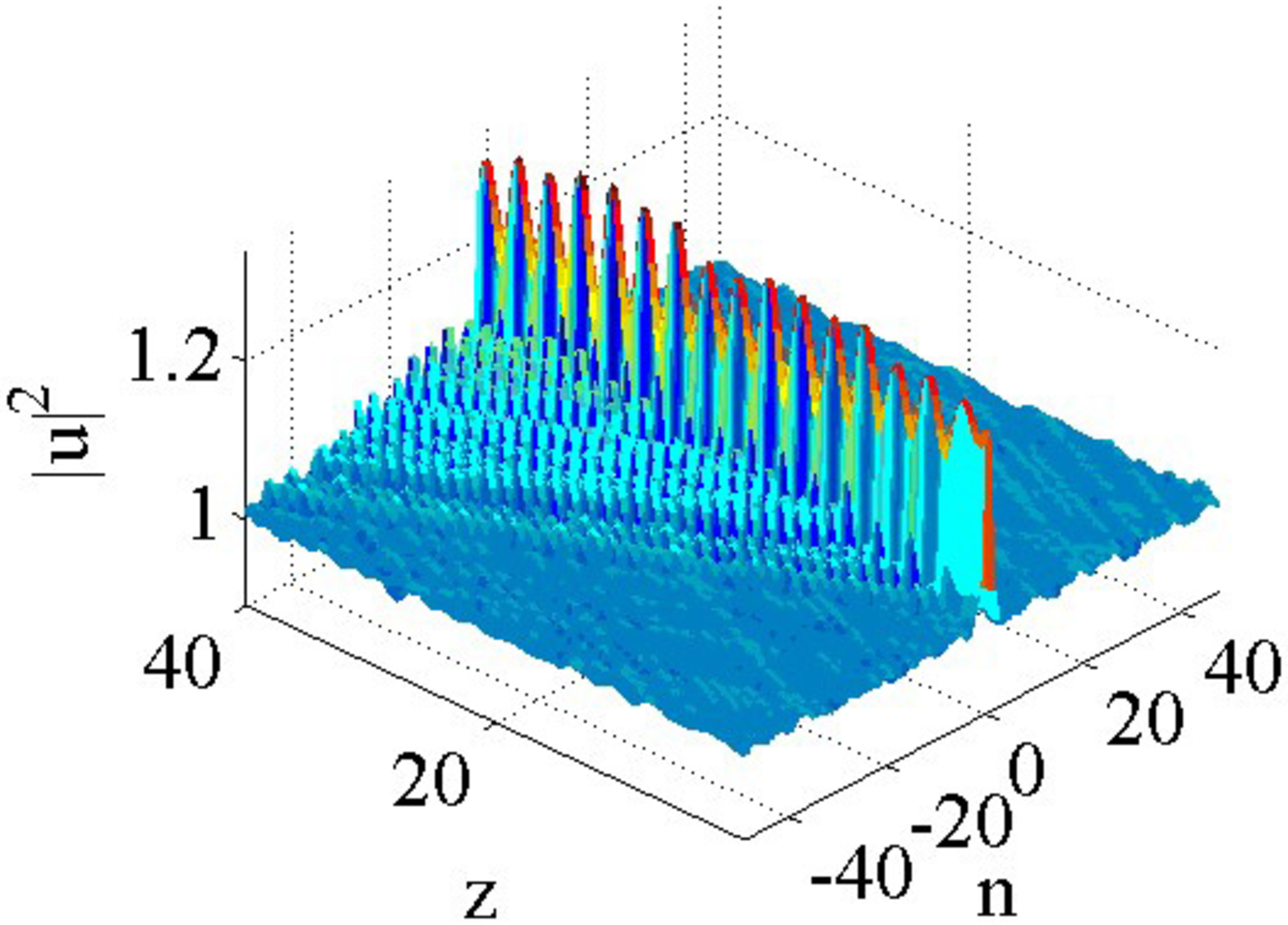}}
\caption{(Color online) The same as in Fig. \protect\ref%
{unstablenonLinearGray}, but for an unstable anti-gray soliton with $\left(
|U_{\mathrm{BG}}|^{2},C_{d}/C_{0},\protect\kappa \right) =(1,2,0.5)$.}
\label{unstablenonLinearAntigray}
\end{figure}

As seen in Fig. \ref{DistributnonLinear}, the increase of the dimer's
intrinsic coupling constant, $C_{d}$, stabilizes the DSs, while the increase
of the gain-loss coefficient, $\kappa $ destabilizes them, as before.
However, the dynamics of unstable gray and anti-gray DSs is different from
the blowup, which was observed for unstable staggered bright DSs [see Fig. %
\ref{Unstablebright}(d)]: as shown in Figs. \ref{unstablenonLinearGray} and %
\ref{unstablenonLinearAntigray}, the instability initiates internal
oscillations in the solitons, and intensive emission of waves propagating on
top of the stable background.

\section{Conclusion}

The objective of this work was to extend the variety of dynamical lattices
with $\mathcal{PT}$-symmetric defects, by introducing the system with the
background defocusing nonlinearity. In addition to the system with the
defect in the form of the dimer with the linear $\mathcal{PT}$ symmetry, a
modification with the nonlinear $\mathcal{PT}$ symmetry was considered too.
The systems can be realized as arrays of optical waveguides with evanescent
coupling. In comparison with the recently introduced model with the $%
\mathcal{PT}$-symmetric dimer embedded into a linear lattice \cite{we}, the
new system gives rise to new types of DSs (discrete solitons), namely
staggered and unstaggered bright ones, and gray and anti-gray DSs, depending
on the relative strength of the dimer's intrinsic coupling constant, $C_{d}$%
. The existence of staggered bright and (unstaggered) gray can be explained
in a qualitative form, with the help of the continuum limit. The boundary
between gray and anti-gray DSs has been predicted too, in an approximate
analytical form. Stability of the modes was investigated through the
computation of the growth rates for small perturbations, and by means of
direct simulations. The existence and stability areas tend to expand with
the increase of $C_{d}$, and shrink with the increase of the gain-loss
coefficient, $\kappa $. In particular, the bright unstaggered modes pinned
to the defect are completely stable. The gray and anti-gray DSs are
completely stable too in the system with the linear $\mathcal{PT}$ symmetry
of the defect, and have a boundary between stable and unstable states in the
case of the nonlinear $\mathcal{PT}$ symmetry. In the latter case, unstable
DSs do not blow up, which is typical for unstable solitons in $\mathcal{PT}$%
-symmetric systems; instead, they develop oscillatory dynamics.

It may be interesting to consider DSs in the self-defocusing lattice with a
pair of defects of the same or opposite signs (dimer-dimer, or
dimer-antidimer), separated by some distance. A challenging perspective for
the extension of the present analysis is to carry it out for the
two-dimensional variant of the models. In that case, the $\mathcal{PT}$%
-symmetric defect may be represented by a dimer or quadrimer, and gray and
anti-gray DSs will be, probably, replaced by vortices.

\begin{acknowledgments}
The authors appreciate useful discussions with Dr. Wei Pang (Guangdong
University of Technology). This work is supported by the National Natural
Science Foundation of China (Grants 11104083 and 11204089). B.A.M.
appreciates hospitality of the Sun Yat Sen University (Guangzhou).
\end{acknowledgments}

%\newpage %Just because of unusual number of tables stacked at end
%

\bibliographystyle{plain}
\bibliography{apssamp}
% Produces the bibliography via BibTeX.

\end{document}